\documentclass[12pt]{article}
\pdfoutput=1

\usepackage{amsmath,amsfonts,enumerate,array}
\usepackage{graphicx}
\usepackage[usenames]{color}
\usepackage[english]{babel}
\usepackage{fancybox}
\usepackage{chngpage} 

\newcommand{\captionfonts}{\small}

\makeatletter  
\long\def\@makecaption#1#2{%
  \vskip\abovecaptionskip
  \sbox\@tempboxa{{\captionfonts #1: #2}}%
 \ifdim \wd\@tempboxa >\hsize
    {\captionfonts #1: #2\par}
  \else
    \hbox to\hsize{\hfil\box\@tempboxa\hfil}%
  \fi
  \vskip\belowcaptionskip}
\makeatother   

\usepackage{mciteplus}

\usepackage[colorlinks=true,      linkcolor=blue,      urlcolor=blue,      filecolor=blue,      citecolor=blue,
      pdfstartview=FitH,     pdfpagemode=UseNone,      bookmarksopen=true]{hyperref}  
      
\usepackage[all]{hypcap}     

\begin{document}

\numberwithin{equation}{section}


\mathchardef\mhyphen="2D


\newcommand{\be}{\begin{equation}}
\newcommand{\ee}{\end{equation}}
\newcommand{\bea}{\begin{eqnarray}\displaystyle}
\newcommand{\eea}{\end{eqnarray}}
\newcommand{\nnm}{\nonumber}
\newcommand{\nn}{\nonumber}
\newcommand{\newotimes}{}  
\newcommand{\utv}{|0_R^{-}\rangle^{(1)}\newotimes |0_R^{-}\rangle^{(2)}}
\newcommand{\utvb}{|\bar 0_R^{-}\rangle^{(1)}\newotimes |\bar 0_R^{-}\rangle^{(2)}}
\newcommand{\diff}{\text{d}}

\def\eq#1{(\ref{#1})}
\newcommand{\secn}[1]{Section~\ref{#1}}

\newcommand{\tbl}[1]{Table~\ref{#1}}
\newcommand{\fig}{Fig.~\ref}

\def\beq{\begin{equation}}
\def\eeq{\end{equation}}
\def\beqa{\begin{eqnarray}}
\def\eeqa{\end{eqnarray}}
\def\bet{\begin{tabular}}
\def\eet{\end{tabular}}
\def\bs{\begin{split}}
\def\es{\end{split}}


\def\a{\alpha}  \def\b\beta{\beta}   \def\c{\chi}    \def\d{\delta}
\def\g{\gamma}  \def\G{\Gamma}  \def\e{\epsilon}  
\def\vep{\varepsilon}   \def\tvep{\widetilde{\varepsilon}}
\def\f{\phi}    \def\F{\Phi}  \def\fb{{\ov \phi}}
\def\vf{\varphi}  \def\m{\mu}  \def\mub{\ov \mu}
\def\n{\nu}  \def\nub{\ov \nu}  \def\o{\omega}
\def\O{\Omega}  \def\r{\rho}  \def\k{\kappa}
\def\kab{\ov \kappa}  \def\s{\sigma}
\def\t{\tau}  \def\th{\theta}  \def\sb{\ov\sigma}  \def\S{\Sigma}
\def\l{\lambda}  \def\L{\Lambda}  \def\p{\psi}

\newcommand{\gt}{\tilde{\gamma}}


\def\cA{{\cal A}} \def\cB{{\cal B}} \def\cC{{\cal C}}
\def\cD{{\cal D}} \def\cE{{\cal E}} \def\cF{{\cal F}}
\def\cG{{\cal G}} \def\cH{{\cal H}} \def\cI{{\cal I}}
\def\cJ{{\cal J}} \def\cK{{\cal K}} \def\cL{{\cal L}}
\def\cM{{\cal M}} \def\cN{{\cal N}} \def\cO{{\cal O}}
\def\cP{{\cal P}} \def\cQ{{\cal Q}} \def\cR{{\cal R}}
\def\cS{{\cal S}} \def\cT{{\cal T}} \def\cU{{\cal U}}
\def\cV{{\cal V}} \def\cW{{\cal W}} \def\cX{{\cal X}}
\def\cY{{\cal Y}} \def\cZ{{\cal Z}}

\def\mC{\mathbb{C}} \def\mP{\mathbb{P}}  
\def\mR{\mathbb{R}} \def\mZ{\mathbb{Z}} 
\def\mT{\mathbb{T}} \def\mN{\mathbb{N}}
\def\mH{\mathbb{H}} \def\mX{\mathbb{X}}

\def\CP{\mathbb{CP}}
\def\RP{\mathbb{RP}}
\def\Z{\mathbb{Z}}
\def\N{\mathbb{N}}
\def\H{\mathbb{H}}

\newcommand{\rmd}{\mathrm{d}}
\newcommand{\rmx}{\mathrm{x}}

\def\tA{ {\widetilde A} } 

\def\one{{\hbox{\kern+.5mm 1\kern-.8mm l}}}
\def\zero{{\hbox{0\kern-1.5mm 0}}}


\newcommand{\bra}[1]{{\langle {#1} |\,}}
\newcommand{\ket}[1]{{\,| {#1} \rangle}}
\newcommand{\braket}[2]{\ensuremath{\langle #1 | #2 \rangle}}
\newcommand{\Braket}[2]{\ensuremath{\langle\, #1 \,|\, #2 \,\rangle}}
\newcommand{\norm}[1]{\ensuremath{\left\| #1 \right\|}}
\def\corr#1{\left\langle \, #1 \, \right\rangle}
\def\vac{|0\rangle}


\def\zb{{\bar z}}

\newcommand{\sq}{\square}
\newcommand{\IP}[2]{\ensuremath{\langle #1 , #2 \rangle}}    

\newcommand{\floor}[1]{\left\lfloor #1 \right\rfloor}
\newcommand{\ceil}[1]{\left\lceil #1 \right\rceil}

\newcommand{\dbyd}[1]{\ensuremath{ \frac{\d}{\d {#1}}}}
\newcommand{\ddbyd}[1]{\ensuremath{ \frac{\d^2}{\d {#1}^2}}}

\newcommand{\Zd}{\ensuremath{ Z^{\dagger}}}
\newcommand{\Xd}{\ensuremath{ X^{\dagger}}}
\newcommand{\Ad}{\ensuremath{ A^{\dagger}}}
\newcommand{\Bd}{\ensuremath{ B^{\dagger}}}
\newcommand{\Ud}{\ensuremath{ U^{\dagger}}}
\newcommand{\Td}{\ensuremath{ T^{\dagger}}}

\newcommand{\T}[3]{\ensuremath{ #1{}^{#2}_{\phantom{#2} \! #3}}}		

\newcommand{\tr}{\operatorname{tr}}
\newcommand{\sech}{\operatorname{sech}}
\newcommand{\Spin}{\operatorname{Spin}}
\newcommand{\Sym}{\operatorname{Sym}}
\newcommand{\Com}{\operatorname{Com}}
\def\adj{\textrm{adj}}
\def\id{\textrm{id}}

\def\ha{\frac{1}{2}}
\def\tha{\tfrac{1}{2}}
\def\wt{\widetilde}
\def\ra{\rangle}
\def\la{\langle}

\def\pb{\ov\psi}
\def\pt{\widetilde{\psi}}
\def\at{\widetilde{\a}}
\def\cb{\ov\chi}
\def\db{\bar\partial}
\def\delb{\bar\partial}
\def\dbar{\ov\partial}
\def\dag{\dagger}
\def\dalpha{{\dot\alpha}}
\def\dbeta{{\dot\beta}}
\def\dgamma{{\dot\gamma}}
\def\ddelta{{\dot\delta}}
\def\ad{{\dot\alpha}}
\def\bd{{\dot\beta}}
\def\dg{{\dot\gamma}}
\def\dd{{\dot\delta}}
\def\th{\theta}
\def\Th{\Theta}
\def\eb{{\ov \epsilon}}
\def\gb{{\ov \gamma}}
\def\wb{{\ov w}}
\def\Wb{{\ov W}}
\def\D{\Delta}
\def\DD{\Delta^\dag}
\def\Db{\ov D}

\def\ov{\overline}
\def\Slash{\, / \! \! \! \!}
\def\dslash{\partial\!\!\!/} 
\def\Dslash{D\!\!\!\!/\,\,}
\def\fslash#1{\slash\!\!\!#1}
\def\Fslash#1{\slash\!\!\!\!#1}

\def\del{\partial}
\def\delb{\bar\partial}
\newcommand{\ex}[1]{{\rm e}^{#1}} 
\def\ii{{i}}

\newcommand{\vs}[1]{\vspace{#1 mm}}

\newcommand{\ve}{{\vec{\e}}}
\newcommand{\shalf}{\frac{1}{2}}

\newcommand{\lb}{\rangle}
\newcommand{\al}{\ensuremath{\alpha'}}
\newcommand{\ap}{\ensuremath{\alpha'}}

\newcommand{\bean}{\begin{eqnarray*}}
\newcommand{\eean}{\end{eqnarray*}}
\newcommand{\ft}[2]{{\textstyle {\frac{#1}{#2}} }}

\newcommand{\hsp}{\hspace{0.5cm}}
\def\half{{\textstyle{1\over2}}}
\let\ci=\cite \let\re=\ref
\let\se=\section \let\sse=\subsection \let\ssse=\subsubsection

\newcommand{\dpb}{D$p$-brane}
\newcommand{\dpbs}{D$p$-branes}

\def\gh{{\rm gh}}
\def\sgh{{\rm sgh}}
\def\NS{{\rm NS}}
\def\R{{\rm R}}
\def\Qp{Q_{\rm P}}
\def\QP{Q_{\rm P}}

\newcommand\dott[2]{#1 \! \cdot \! #2}

\def\eo{\overline{e}}

\newcommand{\ints}{\int\limits}


\def\p{\partial}
\def\h{{1\over 2}}

\def\la{\lambda}
\def\eps{\epsilon}
\def\bb{\bigskip}
\def\tg{\widetilde\gamma}
\newcommand{\dm}{\begin{displaymath}}
\newcommand{\edm}{\end{displaymath}}
\renewcommand{\b}{\widetilde{B}}
\newcommand{\gm}{\Gamma}
\newcommand{\ac}[2]{\ensuremath{\{ #1, #2 \}}}
\renewcommand{\ell}{l}
\newcommand{\z}{\ell}
\def\bb{$\bullet$}
\def\Qbar{{\bar Q}_1}
\def\QPbar{{\bar Q}_p}

\def\q{\quad}

\def\bn{B_\circ}

\let\a=\alpha \let\b=\beta \let\g=\gamma 
\let\e=\epsilon
\let\c=\chi \let\th=\theta  \let\k=\kappa
\let\l=\lambda \let\m=\mu \let\n=\nu \let\x=\xi \let\r=\rho

\let\s=\sigma 
\let\t=\tau 

\let\vp=\varphi \let\vep=\varepsilon
\let\w=\omega  \let\G=\Gamma \let\D=\Delta \let\Th=\Theta \let\P=\Pi \let\S=\Sigma

\def\h{{1\over 2}}

\def\r{\rightarrow}
\def\Ri{\Rightarrow}

\def\nn{\nonumber\\}
\let\bm=\bibitem
\def\Kt{{\widetilde K}}

\let\p=\partial


\begin{flushright}
\end{flushright}

\vspace{15mm}

 \begin{center}
{\LARGE Bogoliubov coefficients for the twist operator \\ \vspace{0.3cm} in the D1D5 CFT}
\\
\vspace{18mm}
{\bf  Zaq Carson\footnote{carson.231@osu.edu}, Samir D. Mathur\footnote{mathur.16@osu.edu} and David Turton\footnote{turton.7@osu.edu} 
\\}
\vspace{12mm}
Department of Physics,\\ The Ohio State University,\\ Columbus,
OH 43210, USA\\ 
\vspace{10mm}
\end{center}

\thispagestyle{empty}
\begin{abstract}

\vspace{2mm}

The D1D5 CFT is a holographic dual of a near-extremal black hole in string theory.  The interaction in this theory involves a twist operator which joins together different copies of a free CFT.  Given a large number of D1 and D5 branes, the effective length of the circle on which the CFT lives is very large.  We develop a technique to study the effect of the twist operator in the limit where the wavelengths of excitations are short compared to this effective length, which we call the `continuum limit'.  The method uses Bogoliubov coefficients to compute the effect of the twist operator in this limit.  For bosonic fields, we use the method to reproduce recent results describing the effect of the twist operator when it links together CFT copies with windings $M$ and $N$, producing a copy of winding $M+N$.  We also comment on possible generalizations of our results.  The methods developed here may help in understanding the twist interaction at higher orders.  This in turn should provide insight into the thermalization process in the D1D5 CFT, which gives a holographic description of black hole formation.
\end{abstract}
\newpage

\section{Introduction}
\label{intr}\setcounter{footnote}{0}

\baselineskip=15pt
\parskip=3pt

It would be very interesting if we could have an understanding of the detailed dynamics of black hole formation. In string theory we have a concrete description of a near-extremal black hole, given by the D1D5 system \cite{sv,callanmalda}.  By the AdS/CFT correspondence \cite{adscft}, we can hope to learn about the physics of black holes by studying the physics of the CFT which describes the low energy dynamics of the D1D5 bound state.

It is easiest to study the  D1D5 CFT at its `free' point, termed the `orbifold point' in coupling space \cite{orbifold}.  The gravity solution describing the black hole does not correspond to this free point; one must add an appropriate deformation operator to the free Hamiltonian. In spite of this fact, one finds that many properties of the black hole such as the rate of Hawking radiation and greybody factors are reproduced by the free theory  \cite{dasmathur,maldastrom}. 

The process of black hole {\it formation}, however, cannot be understood by using the free theory. This process is dual to a thermalization process in the dual CFT, and the free theory does not have an interaction that can lead to thermalization. We are thus led to study the deformation operator, which plays the role of the interaction in the D1D5 CFT. 

In \cite{acm2,acm3,chmt}, the effect of this deformation operator on simple initial states was obtained. However the result is somewhat complicated, involving ratios of gamma functions. Importantly, it appears that the exact methods of \cite{acm2,acm3,chmt} cannot be extended to more general computations, such as the insertion of multiple twist operators; the complexity of the calculations prevent one from obtaining answers in closed form.

Fortunately, the physics of black holes involves a large number of D1 and D5 branes, and in this situation the `effective length' of the circle on which the CFT lives is very large \cite{maldasuss}. Thus it is useful to develop an approach where one can directly obtain the effect of the twist operator in the limit where the wavelength of the excitations is short compared to the length of the effective CFT circle. We  call this limit the `continuum limit', since the mode numbers $k$ of excitations $a^\dagger_k$ go from being discrete to being `almost continuous'.  It was noted in \cite{chmt} that the results of that paper simplify considerably in this  limit.

The goal of this paper is to develop a method of obtaining the results of \cite{acm2,acm3,chmt} directly in the continuum limit, with the hope that this method will be extendable to more general computations. In particular, it is hoped that this new method can be extended to the insertion of multiple twist operators, thus allowing us to consider  several orders in perturbation theory around the orbifold point.  Our method uses Bogoliubov coefficients to compute the effect of the twist operator in the continuum limit.  The approach is somewhat similar to that of \cite{ac}, but is different in the details.  We will comment further on this in due course.

We next discuss in more detail the twist operator, the quantities we compute, and the Bogoliubov method.

\begin{figure}[t]
\begin{center}
\includegraphics[scale=.50]{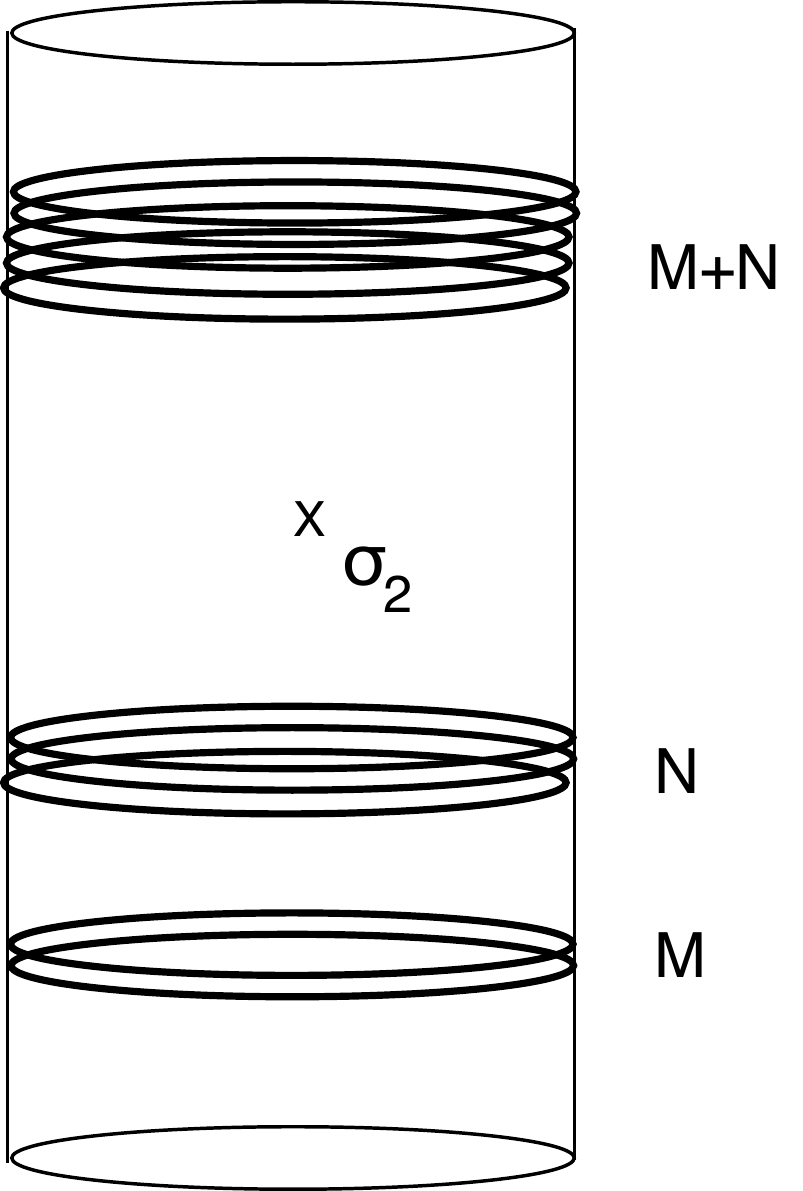}
\caption{The cylinder with coordinate $w$. The state before the twist has component strings with windings $M,N$. The twist operator $\sigma_2$  links these into a single component string of winding $M+N$.}
\label{fone}
\end{center}
\end{figure}

\subsection{The twist operator}

Let the D1D5 bound state be composed of $N_1$ D1 branes and $N_5$ D5 branes.   At the orbifold point the CFT is described by a symmetric product of $N_1N_5$ copies of a free CFT, where each copy contains 4 bosons and 4 fermions. Since the different copies are symmetrized, the operator content includes twist operators. A twist operator $\sigma_n$ takes $n$ different copies of the CFT and links them into a single copy living on a circle that is $n$ times longer. We call any such linked set of copies a `component string'. 

The deformation operator  also has the form of a twist operator $\sigma_2$, dressed with a supercharge: $\hat O\sim G_{-\h} \sigma_2$. The effect of the twist is depicted in Fig.\;\ref{fone}. Before the interaction,  we have component strings with windings $M, N$. The interaction links these component strings together, generating a component string with winding $M+N$.\footnote{In the present paper we focus on the process of two component strings being joined together. If the twist operator acts on two strands of the same component string, then it will cut the component string into two parts. The computations for that case can be done in a similar way to the computations presented here.}  The twist operator does not affect the flavor indices of the bosons, so we will suppress these indices throughout this paper.

There are two relevant effects of this twist:

(a) Suppose the CFT on both circles in Fig.\;\ref{fone}(a) is in the vacuum state; thus the state is $|0^{(1)}\rangle |0^{(2)}\rangle$, where the superscripts differentiate between the two component strings. After the twist, the CFT will {\it not} be in the vacuum state $|0\rangle$ of the CFT on the component string with winding $M + N$. Denoting the canonically normalized bosonic modes on the component string of length $M+N$ as $a_s$, the state will be of the form:
\be
|0^{(1)}\rangle |0^{(2)}\rangle ~~\r ~~
|\chi\rangle~~\sim~~ e^{\sum_{s,s'}\gamma^B_{ss'}a^{\dagger}_{s}a^{\dagger}_{s'}}|0\rangle
\ee
The coefficients $\gamma^B_{ss'}$ were found in \cite{acm2} for the case where the initial windings were $M=N=1$, and in \cite{chmt} for the case of general $M,N$. The method in each case was to pass to a `covering space' where the effect of all the twists was undone, and the computations were reduced to those in a free CFT on the sphere. But such a covering map is not in general easy to find or compute with if we have many twist insertions.

(b) Let the canonically normalized modes on the component string with winding $M$ be given by $ a^{(1)}_q$, and let the canonically normalized modes on the component string with winding $N$ be given by $a^{(2)}_r$.  Suppose we start with an initial excitation $a^{(1)\dagger}_{q}$ on the component string with winding $M$ before the twist. After the twist, this excitation gets converted to a linear combination of excitations above the state $|\chi\rangle$,
\be
a^{(1)\dagger}_{q} |0^{(1)}\rangle |0^{(2)}\rangle ~~\r ~~\sum\limits_s f^{B(1)}_{qs}\,  a^{\dagger}_{s}|\chi\rangle \,, \qquad i=1,2
\ee
on the final component string of length  $M+N$. A similar relation holds for an initial excitation $a^{(2)\dagger}_r$ on the component string of winding $N$, defining coefficients $f^{B(2)}_{rs}$.  The coefficients $f^{B(1)}_{qs}$, $f^{B(2)}_{rs}$ were found  in \cite{acm3} for the case where the initial windings were $M=N=1$, and in \cite{chmt} for the case of general $M,N$. Again the method involves passing to a covering space, and this becomes complicated if there are many twist insertions.

\subsection{The computations we perform}

We will restrict ourselves to bosonic excitations in the present paper. Fermions are expected to behave in a similar manner, but involve careful consideration of the spin carried by the vacuum state; thus we postpone their discussion to a separate study.

\begin{figure}[t]
\begin{center}
\includegraphics[scale=.55]{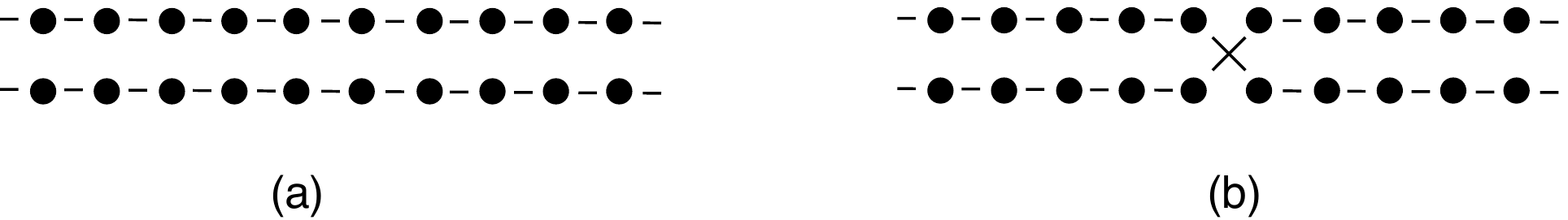}
\caption{(a) The scalar field on the component strings modeled by point masses joined by springs. (b) The twist operator $\sigma_2$ changes the springs so that the masses are linked in a different way.}
\label{ftwo}
\end{center}
\end{figure}

The twist operator takes a free theory (the CFT living on two separate circles of lengths  $2\pi M,2\pi N$) to another  free theory (the CFT living on a circle of length $2\pi(M+N)$). To understand the effect of such a twist, consider a discretization of a 1+1 dimensional bosonic free field $X$.  We can model a free bosonic  field by a collection of point masses joined by springs. This gives a set of coupled harmonic oscillators, and the oscillation amplitude of the masses then gives the field $X(\tau,  \sigma)$. Consider such a collection of point masses on two different circles, and let the state in each case be the ground state of the coupled oscillators (Fig.\;\ref{ftwo}(a)). At time $\tau_0$ and position $\sigma_0$, we insert a twist $\sigma_2$. The effect of this twist is to connect the masses with a different set of springs, so that the masses make a single chain of longer  length (Fig.\;\ref{ftwo}(b)).

Recall that we have modes $a^{(1)}_{q}$ on the component string with winding $M$, and modes $a^{(2)}_{r}$ on the component string with winding $N$.  On application of the twist, the field $X(\s,\t)$ does not immediately change, but the new couplings imply that we should expand this field in terms the of oscillator modes $a_s$ on the twisted string. There is a linear
relation between oscillators before and after the twist, involving both left and right movers:
\bea
a^{(1)}_q&=&\alpha^{(1)}_{qs}a_s+ \alpha^{(1)}_{q\bar s}{\bar a}_{\bar s}+\beta^{(1)}_{qs} a^\dagger_s+\beta^{(1)}_{q\bar s} {\bar a}^\dagger_{\bar s}\nn
a^{(2)}_r&=&\alpha^{(2)}_{rs}a_s+ \alpha^{(2)}_{r\bar s}{\bar a}_{\bar s}+\beta^{(2)}_{rs} a^\dagger_s+\beta^{(2)}_{r\bar s} {\bar a}^\dagger_{\bar s}
\eea

We collect $\alpha^{(1)}, \alpha^{(2)}$ into a single matrix $\alpha$ relating the annihilation operators before the twist to  the annihilation operators after the twist, and similarly we collect $\beta^{(1)}, \beta^{(2)}$ into a single matrix $\beta$ relating the creation operators before the twist to  the annihilation operators after the twist. We recall the well-known fact that the matrix $\gamma^B$ is given by
\be
\gamma^B=\alpha^{-1}\beta \,.
\ee
We can also group the functions $f^{B(1)}_{qs}, f^{B(2)}_{rs}$ into a matrix $f^B$ relating the excitation before the twist to the excitations after the twist. We will note that the following general relation holds:
\be
f^B=\left ( \alpha^{-1}\right )^T \,.
\ee

In order to calculate the physical quantities $\g^B$ and $f^B$, we must first compute the matrices $\alpha, \beta$. This computation is quite simple, since the elements of these matrices are just given by inner products of mode functions before and after the twist. However, the computations of $\gamma^B$ and $f^B$ also require the inverse $\alpha^{-1}$.  The matrix $\a$ is infinite, and we will come across some subtleties regarding uniqueness of inverses and associativity of multiplication.  We will comment on such issues in due course.  Here we simply note that:  

\begin{enumerate}[(a)]
\item The quantity $f^B = \left ( \a^{-1} \right )^T$ is a unique, well-defined physical quantity.
\item We can discretize our problem as described above with point masses and springs.  If we do this, $\a$ becomes a finite matrix with a unique inverse. 
\end{enumerate}

It is not a priori clear how to go about computing $\a^{-1}$.  One approach is to make an ansatz for $\a^{-1}$ in the continuum limit, and then verify that it satisfies the required properties.  This is the approach we will follow in the current paper.  Having made the ansatz, the major effort of this paper goes to verifying that it is correct.\footnote{After this work was substantially complete, exact expressions for $\g^B$ and $f^B = \left ( \a^{-1} \right )^T$ were obtained by independent methods \cite{chmt}.}  The check $\alpha\alpha^{-1}=\alpha^{-1}\a=\one$ involves multiplying infinite matrices; we carry out this multiplication by using approximations appropriate to the continuum limit, which in several places allows us to replace index sums by integrals. These integrals are then evaluated by contour methods.

Let us now make a comment regarding previous work.  In \cite{ac}, a relation was written between operators before and after the twist, but the corresponding matrices $\alpha, \beta$ were different from the ones we compute in the present paper.  Nevertheless we will see that both approaches lead to the same physical quantities $\a^{-1}$ and $\g^B$.  We thus find that as an infinite matrix, the physical quantity $\a^{-1}$ does not have a unique inverse. This leads to a one-parameter family of different but equivalent choices for $\a$ and $\b$.  We will explore this in Appendix \ref{NonUnique}.

The deformed D1D5 CFT has recently been studied in various other works. The effect of the twist operator on entanglement entropy was studied in \cite{aa}. Twist-nontwist correlators were calculated in \cite{peet1}, and operator mixing was investigated in \cite{peet2}. For other related work, see \cite{orbifold2}. The present line of enquiry complements the fuzzball program~\cite{fuzzball}; for recent work in this area, see e.g.~\cite{fuzzrecent1,fuzzrecent2}.

The remainder of this paper is organized as follows.  In Section \ref{matsec:2} we review the Bogoliubov formalism in the context of particle creation in curved space.   In Section \ref{1,1twist} we calculate $\a,\b,$ and $\a^{-1}$ for the case $M=N=1$.  In Section \ref{1,1gamma} we calculate the quantity $\g^B$ for $M=N=1$.  In Section \ref{generalMN} we present $\a, \b,$ and $\a^{-1}$ for general $M$ and $N$.  In Section \ref{gammaMN} we compute $\g^B$ for general $M$ and $N$.  In Section \ref{Discuss} we discuss our results.  Various technical details are presented in the appendices.

\section{Particle creation in curved space}
\label{matsec:2}

We will use the formalism of Bogoliubov coefficients to compute the effect of the twist operator. We begin by first reviewing the formalism in the context of quantum fields in curved space.  We will then show how it can be adapted for our problem. 

Consider a free scalar field $\phi$ in curved spacetime, satisfying $\square \phi=0$.  Given a complete Cauchy hypersurface $\Sigma$ and smooth complex functions $f$ and $g$, we have the inner product:
\begin{equation}
(f, g)\equiv -i\ints_\Sigma d\Sigma^\mu \left ( f\partial_\mu g^*-g^*\partial_\mu f \right )
\label{inner}
\end{equation}
There are many ways in which one can choose the time coordinate in curved space. For one such choice, let the positive frequency solutions of the wave-equation be $f_q(x)$; then their complex conjugates give negative frequency modes $f_q^*(x)$. We require that the $f_q$ form a complete orthonormal set of solutions with respect to the above inner product,
\begin{equation}
(f_q, f_{q'})=\delta_{qq'}, ~~~(f_q, f^*_{q'})=0, ~~~(f^*_q, f^*_{q'})=-\delta_{qq'}
\end{equation}
We then expand the field operator $\phi$ as
\begin{equation}
\phi(x)=\sum_q \left (a_q f_q(x)+a^\dagger_q f_q^*(x)\right  )
\end{equation}
We  define the $a$-vacuum $|0\rangle_a$ to be the state which is annihilated by all the $a_q$ annihilation operators:
\begin{equation}
a_q|0\rangle_a=0
\end{equation}

Now consider a different time coordinate; with this coordinate, let $h_s(x)$ be a complete orthonormal set of positive frequency modes.  We then have the expansion
\begin{equation}
\phi(x)=\sum_s \left (b_s h_s(x)+b^\dagger_s h_s^*(x)\right  )
\end{equation}
Similarly, the $b$-vacuum $|0\rangle_b$ is defined as the state which is annihilated by all the $b_s$ annihilation operators:
\begin{equation}
b_s |0\rangle_b=0
\end{equation}
Now from the two different expansions of $\phi$ we have
\begin{equation} \label{eq:2exp}
\sum_q \left (a_q f_q(x)+a^\dagger_q f_q^*(x)\right  )
=\sum_s \left (b_s h_s(x)+b^\dagger_s h_s^*(x)\right  )
\end{equation}

We now define the Bogoliubov coefficients $\a$ and $\b$ as follows:
\begin{equation} \label{jone}
a_q\equiv \sum_s \alpha_{qs} b_s+\sum_s \beta_{qs} b^\dagger_s \,.
\end{equation}
Taking the inner product with $f_q$ on each side of \eqref{eq:2exp}, we obtain
\bea
\alpha_{qs} = (h_s, f_q) \,, \qquad \beta_{qs} = (h^*_s, f_q)  \,.
\eea
From the fact that 
\be
[a_q, a^\dagger_{q'}]=\delta_{qq'}, ~~~~~~[b_s, b^\dagger_{s'}]=\delta_{ss'}
\ee
we find that
\be \label{eq:adbd}
\alpha\alpha^\dagger - \beta \beta^\dagger=\one
\ee
We also see that the vacuum $|0\rangle_a$ satisfies
\begin{equation}
0=a_q|0\rangle_a=\left (\sum_s \alpha_{qs} b_s+\sum_s \beta_{qs} b^\dagger_s\right )|0\rangle_a
\label{matqtwo}
\end{equation}
This relation has the solution 
\begin{equation}
|0\rangle_a=C e^{-{1\over 2}\sum_{s,s'}b_s^\dagger\gamma^B_{ss'} b^\dagger_{s'}} |0\rangle_b
\label{matdone}
\end{equation}
where the matrix $\gamma^B$ is given by 
\begin{equation}
\gamma^B= \alpha^{-1} \beta
\end{equation}
We also note that $\g^B$ is symmetric.

\subsection{The relation $f^B=\left (\alpha^{-1}\right )^T$}

Now consider a state that starts with an initial $a$ excitation,
\be
a^\dagger_q |0\rangle_a \,.
\ee
From (\ref{jone}) we have the relation 
\be
a^\dagger=\alpha^* b^\dagger +\beta^* b
\ee
Thus we have
\bea
a^\dagger_q |0\rangle_a &=& C(\alpha^* b^\dagger +\beta^* b)_q e^{-\h b^\dagger \gamma^B b^\dagger}|0\rangle_b
~=~C[(\alpha^*  -\beta^* \gamma)b^\dagger]_q e^{-\h b^\dagger \gamma^B b^\dagger}|0\rangle_b \,.
\eea
We define
\be \label{eq:fbdef}
f^B\equiv \alpha^*  -\beta^* \gamma^B
\ee
and so in terms of the coefficients $f^B_{qs}$, we have:
\be \label{eq:fdef}
a^\dagger_q |0\rangle_a = C\sum_s f^B_{qs} \; b^\dagger_s \; e^{-\h b^\dagger \gamma^B b^\dagger}|0\rangle_b \,.
\ee
Now, we note that
\be
a_q a^\dagger_{q'} |0\rangle_a = \delta_{qq'}|0\rangle_a \,.
\ee
The LHS of this equation may be expanded using \eq{jone} and \eq{eq:fdef}, giving
\be
C\left(\alpha_{qs} b_s +\beta_{qs} b^\dagger_s\right)\left ( f^B_{q's'} \, b^\dagger_{s'} \right ) e^{-\h b^\dagger \gamma^B b^\dagger}|0\rangle_b 
\ee
Since $\alpha \gamma = \beta$, we find a cancellation between the $\b$ term and the term where $b_{s}$ contracts with the $b^{\dagger}$ modes in the exponent.  This leaves only the term where $b_{s}$ contracts with $b^\dagger_{s'}$, giving
\be
\left[(\alpha \left (f^B\right )^T\right]_{qq'}Ce^{-\h b^\dagger \gamma^B b^\dagger}|0\rangle_b = \delta_{qq'}|0\rangle_a \;.
\ee
Thus we see that consistency requires the general relation
\be
\alpha \left (f^B\right )^T=\one \,.
\ee
If we assume that the Bogoliubov matrices may be multiplied associatively, we can also show in general that $\left (f^B\right )^T \alpha =\one$:
\bea
\left(\alpha^*  -\beta^* \gamma^B\right)^T \a &=&
\a^\dag \a - \left( \a^{-1} \b \right) \b^{\dag} \a \nn
&=& \a^\dag \a - \a^{-1} \left( \b  \b^{\dag} \right) \a \\
&=& \a^\dag \a - \a^{-1} \left( - \one + \a \a^\dag \right) \a ~=~ \one \,\nonumber
\eea
where we used \eq{eq:adbd}. In our setup, we will however encounter non-associativity; this will be discussed in Appendix \ref{NonUnique}.  We will separately verify that in our setup we indeed have $\left (f^B\right )^T\a = \one$, and so we write:
\be
f^B = \left ( \a^{-1}\right )^T.
\ee

\section{The twist $1+1\r 2$}\label{1,1twist}
\subsection{Mode expansions on the cylinder}
Let us now turn to the computation of Bogoliubov coefficients in the case of the twist insertion. Consider a scalar field $X(\sigma, \tau)$ in the 1+1 dimensional CFT. Let the situation before the twist be the one depicted in Fig.\;\ref{fone}(a) where we have two separate circles, each wound only once.  We then have two such fields, $X^{(1)}$ and $X^{(2)}$. We can expand $X^{(1)}$ and $X^{(2)}$ in terms of mode functions on the circles.\footnote{Throughout this paper, except where explicitly stated, we will ignore zero-modes.}  These modes come in both left-moving and right-moving forms.  We use a bar to denote right-moving modes.  Before the twist, we have the expansions:
\bea
X^{(1)}(\sigma, \tau)&=&\sum_{q>0} \left ( a^{(1)}_q f^{(1)}_q(\sigma, \tau) + a^{(1)\dagger}_q f^{(1)*}_q(\sigma, \tau)\right )\nn
&&{}+\sum_{\bar q>0} \left ({\bar a}^{(1)}_{\bar q} f^{(1)}_{\bar q}(\sigma, \tau) + {\bar a}^{(1) \dagger}_{\bar q} f^{(1)*}_{\bar q}(\sigma_, \tau)\right )
\eea
and
\bea
X^{(2)}(\sigma, \tau)&=&\sum_{r>0} \left ( a^{(2)}_r f^{(2)}_r(\sigma, \tau) + a^{(2)\dagger}_r f^{(2)*}_r(\sigma, \tau)\right )\nn
&&{}+\sum_{\bar r>0} \left ({\bar a}^{(2)}_{\bar r} f^{(2)}_{\bar r}(\sigma, \tau) + {\bar a}^{(2) \dagger}_{\bar r} f^{(2)*}_{\bar r}(\sigma, \tau)\right )
\eea
 where the superscripts $1,2$ refer to the two circles, and $0 \leq \s < 2\pi$.  The mode functions are given by
\bea
f^{(1)}_q&=&{1\over \sqrt{2\pi}} {1\over \sqrt{2q} }e^{iq(\sigma-\tau)}, ~~~q\in \mathbb{Z},~~q>0, ~~~0\le \sigma<2\pi \\
f^{(1)}_{\bar q} & = & {1\over \sqrt{2\pi}} {1\over \sqrt{2 \bar q} }e^{-i \bar q (\sigma+\tau)}, ~~~\bar q\in \mathbb{Z}, ~~ \bar q >0, ~~~0\le \sigma<2\pi \\
f^{(2)}_r&=&{1\over \sqrt{2\pi}} {1\over \sqrt{2r} }e^{ir(\sigma-\tau)},  ~~~r \in \mathbb{Z},~~r >0, ~~~0\le \sigma<2\pi \\
f^{(2)}_{\bar r}&=&{1\over \sqrt{2\pi}} {1\over \sqrt{2 \bar r} }e^{-i \bar r (\sigma+\tau)},  ~~~\bar r \in \mathbb{Z}, ~~ \bar r >0, ~~~0\le \sigma<2\pi
\eea

 Now we apply the twist operator at the point $(\sigma_0, \tau_0)$, converting the two singly wound strings to one doubly wound string. Then we have a single field $X$, with expansion:
\bea
X(\sigma, \tau)&=&\sum_{s>0} \left (a_s f_s(\sigma, \tau) + a^{\dagger}_sf^*_s(\sigma, \tau)\right ) + \sum_{\bar s>0} \left ({\bar a}_{\bar s} f_{\bar s}(\sigma, \tau) + {\bar a}^{\dagger}_{\bar s} f^{*}_{\bar s}(\sigma, \tau)\right )
 \eea
 where now $0 \leq \s < 4\pi$.  The modes on the final (doubly wound) component string are
\bea
f_s&=&{1\over \sqrt{4\pi}}{1\over \sqrt{2s}}e^{is(\sigma-\tau)},  ~~~s\in \h\mathbb{Z},~~s>0, ~~~0\le \sigma < 4\pi \\
f_{\bar s}&=&{1\over \sqrt{4\pi}}{1\over \sqrt{2 \bar s}}e^{-i \bar s(\sigma+\tau)},  ~~~\bar s\in \h\mathbb{Z},~~\bar s >0, ~~~0\le \sigma < 4\pi.
\eea

At the time of the twist insertion, the fields are related in a straightforward fashion.  We choose a starting spatial position $\s = 0$ on the cylinder.  At this point, we set the field $X$ on the final component string to be equal to the field $X^{(1)}$.  We then move around the cylinder until we reach $\s = \s_0$.  Here the twist moves us to the second copy of the initial component strings, and we have $X = X^{(2)}$.  We continue moving around the cylinder until we return to the twist insertion with $\s = \s_0 + 2\pi$.  Here the twist moves us back to first copy of the initial component strings, so we again have $X = X^{(1)}$.  This gives the relations:
\bea
X(0 \leq \s < \s_0, \t_0) & = & X^{(1)}(\s,\t_0) \nn
X(\s_0 \leq \s < 2\pi + \s_0, \t_0) & = & X^{(2)}(\s,\t_0) \nn
X(2\pi + \s_0 \leq \s < 4\pi, \t_0) & = & X^{(1)}(\s,\t_0)
\eea
where we have used the fact that $X^{(1)}$ and $X^{(2)}$ are both $2\pi$ periodic.

\subsection{Computing the Bogoliubov coefficients}
The wavefunctional of the field $X$ does not change at the instant the twist is applied. We can express this wavefunctional either in terms of the modes before the twist or in terms of the modes after the twist. Suppose $X$ is in the vacuum state $|0^{(1)}\rangle  |0^{(2)}\rangle$ before the twist. After the twist,  we will not have the vacuum state of the twisted string. To find the state, we must calculate the matrix $\g^B$.  But first we will have to calculate $\a$, $\b$, and $\a^{-1}$.

The matrices $\a$ and $\b$ are defined in terms of the inner product (\ref{inner}).  In our case, the surface $\Sigma$ is the entirety of the doubly-wound component string at $\tau = \t_0$.  But since the mode functions $f^{(1)}$ and $f^{(2)}$ do not individually span the full $4\pi$ range of the integration, only certain regions will have a nonzero integrand for any given $f^{(1)}$ or $f^{(2)}$.  The inner product (\ref{inner}) then takes the form:
\bea\label{innerproduct}
(f_s,f^{(1)}_q)&=&-i\ints_{0}^{\s_0} d\sigma \left (f_s\,  \p_\tau f^{(1)*}_q - f^{(1)*}_q \p_\tau f_s \right ) -i\ints_{2\pi + \s_0}^{4\pi} d\sigma \left (f_s\,  \p_\tau f^{(1)*}_q - f^{(1)*}_q\p_\tau f_s \right ) \nn
(f_s,f^{(2)}_r)&=&-i\ints_{\s_0}^{2\pi + \s_0} d\sigma \left (f_s\,  \p_\tau f^{(2)*}_r - f^{(2)*}_r\p_\tau f_s\right )
\eea
and similarly for the anti-holomorphic mode functions.

The Bogoliubov coefficients $\a^{(1)}_{qs}$, $\b^{(1)}_{qs},$ etc.~are defined as follows:
\bea
a^{(1)}_q&=&\alpha^{(1)}_{qs}a_s+ \alpha^{(1)}_{q\bar s}{\bar a}_{\bar s}+\beta^{(1)}_{qs} a^\dagger_s+\beta^{(1)}_{q\bar s} {\bar a}^\dagger_{\bar s}\nn
a^{(2)}_r&=&\alpha^{(2)}_{rs}a_s+ \alpha^{(2)}_{r\bar s}{\bar a}_{\bar s}+\beta^{(2)}_{rs} a^\dagger_s+\beta^{(2)}_{r\bar s} {\bar a}^\dagger_{\bar s}
\eea
Using the inner product in the form of (\ref{innerproduct}), we find
\bea
\alpha^{(1)}_{qs}&=&\left ( f_s,f^{(1)}_q \right ) ~=~ \begin{cases}
{1\over \sqrt{2}}\delta_{qs} & s \in \mathbb{Z} \\
{1 \over 2\pi i \sqrt{2}}{s+q \over \sqrt{sq}}{1\over s-q}e^{i(s-q)(\sigma_0-\tau_0)} & s \in \mathbb{Z}+\h
\end{cases}\nn
\a^{(1)}_{q\bar s}&=&\left ( f_{\bar s},f^{(1)}_q \right ) ~=~ \begin{cases}
0 & \bar s \in \mathbb{Z} \\
-{1 \over 2\pi i \sqrt{2}}{1 \over \sqrt{\bar s q}}e^{-i(\bar s + q)\s_0}e^{-i(\bar s - q)\t_0} & \bar s \in \mathbb{Z} + \h
\end{cases}
\eea
\bea\label{beta1}
\beta^{(1)}_{qs}&=&\left ( f_s^*,f^{(1)}_q \right ) ~=~ \begin{cases}
0 & s \in \mathbb{Z} \\
{1 \over 2\pi i \sqrt{2}}{s-q \over \sqrt{sq}}{1\over s+q}e^{-i(s+q)(\sigma_0-\tau_0)} & s \in \mathbb{Z} + \h
\end{cases}\nn
\beta^{(1)}_{q\bar s}&=&\left ( f_{\bar s}^*,f^{(1)}_q \right ) ~=~ \begin{cases}
0 & \bar s \in \mathbb{Z} \\
-{1 \over 2\pi i \sqrt{2}}{1 \over \sqrt{\bar s q}}e^{i(\bar s - q)\s_0}e^{i(\bar s + q)\t_0} & \bar s \in \mathbb{Z} + \h
\end{cases}
\eea
and
\bea
\alpha^{(2)}_{rs}&=& \left ( f_s,f^{(2)}_r \right ) ~=~ \begin{cases}
{1\over \sqrt{2}}\delta_{rs} & s\in \mathbb{Z} \\
-{1 \over 2\pi i \sqrt{2}}{s+r \over \sqrt{sr}}{1\over s-r}e^{i(s-r)(\sigma_0-\tau_0)} & s\in \mathbb{Z}+\h
\end{cases}\nn
\a^{(2)}_{r\bar s}&=&\left ( f_{\bar s},f^{(2)}_r \right ) ~=~ \begin{cases}
0 & \bar s \in \mathbb{Z} \\
{1 \over 2\pi i \sqrt{2}}{1 \over \sqrt{\bar s r}}e^{-i(\bar s + r)\s_0}e^{-i(\bar s - r)\t_0} & \bar s \in \mathbb{Z} + \h
\end{cases}
\eea
\bea\label{beta2}
\beta^{(2)}_{rs}&=&\left ( f_s^*,f^{(2)}_r \right ) ~=~ \begin{cases}
0 & s \in \mathbb{Z} \\
-{1 \over 2\pi i \sqrt{2}}{s-r \over \sqrt{sr}}{1\over s+r}e^{-i(s+r)(\sigma_0-\tau_0)}  & s \in \mathbb{Z} + \h
\end{cases}\nn
\b^{(2)}_{r\bar s}&=&\left ( f_{\bar s}^*,f^{(2)}_r \right ) ~=~ \begin{cases}
0 & \bar s \in \mathbb{Z} \\
{1 \over 2\pi i \sqrt{2}}{1 \over \sqrt{\bar s r}}e^{i(\bar s - r)\s_0}e^{i(\bar s + r)\t_0} & \bar s \in \mathbb{Z} + \h
\end{cases}
\eea

For later convenience we introduce the notation
\be
s=s_0+{s_2\over 2}
\ee
where $s_0\in \mathbb{Z}$ and $s_2$ can be $0$ or $1$. Thus the two possible values of $s_2$ correspond to $s\in \mathbb{Z}$ and $s\in \mathbb{Z}+\h$ respectively.  We use an analogous notation for $\bar s$.

\subsection{Finding $\alpha^{-1}$}

So far, our calculations have been exact.  We now wish to find an approximation for $\a^{-1}$ in the `continuum limit', as follows. We consider $M$ and $N$  to be some given fixed positive integers, and we take all momentum quantities to be of order some momentum scale $k$ which is large compared to the lowest mode on each component string:
\be
k \gg {1\over M}\,, \quad k \gg {1 \over N} \qquad \Rightarrow \qquad k \gg {1\over M+N}\,.
\ee
We will carry out our computations to leading order in an expansion in negative powers of $Mk$ $(\sim Nk)$. In this section we have $M = N = 1$, and thus $k \gg 1$.

To leading order in our approximation, the matrix $\a^{-1}$ should satisfy the relation:
\be\label{inverserelation}
\a \a^{-1} = \a^{-1}\a = \one \,.
\ee
For ease of notation, we will mostly write equalities as done here,  and leave implicit the fact that we work to order $k^{0}$.

The non-zero index combinations of the relation (\ref{inverserelation}) are:
\be
\sum_{s}\alpha^{(1)} _{qs}\left (\alpha^{-1}\right)^{(1)} _{sq'} =  \d_{qq'}, \qquad \sum_{s}\alpha^{(2)} _{rs}\left (\alpha^{-1}\right)^{(2)} _{sr'} =  \d_{rr'} \label{keyrelation1}
\ee
and
\be
\sum_{q}\left (\alpha^{-1}\right)^{(1)} _{sq}\alpha^{(1)} _{qs'}+\sum_{r}\left (\alpha^{-1}\right)^{(2)} _{sr}\alpha^{(2)} _{rs'}  =  \d_{ss'} \label{leftinverse}
\ee
and similarly for the purely anti-holomorphic combinations.  All other combinations are required to vanish, for example:
\be
\sum_{s}\alpha^{(1)} _{qs}\left (\alpha^{-1}\right)^{(2)} _{sr} =  \sum_{s}\alpha^{(2)} _{rs}\left (\alpha^{-1}\right)^{(1)} _{sq} =  0\label{keyrelation2}
\ee
and so on.

Our strategy is to attempt a simple ansatz, and check that it satisfies the above relations.  Let us begin with the ansatz
\bea
\left (\alpha^{-1}\right)^{(1)} _{sq}&=& \begin{cases}
C_{(1)}\delta_{sq} & s \in \mathbb{Z} \\
C'_{(1)}{1\over s-q} \, e^{-i(s-q)(\sigma_0-\tau_0)} & s \in \mathbb{Z}+\h
\end{cases}\nn
\left (\alpha^{-1}\right)^{(1)} _{\bar s q}&=&0
\eea
and
\bea
\left (\alpha^{-1}\right)^{(2)} _{sr}&=& \begin{cases}
C_{(2)}\delta_{sr} & s \in \mathbb{Z} \\
C'_{(2)}{1\over s-r} \, e^{-i(s-r)(\sigma_0-\tau_0)} & s \in \mathbb{Z}+\h
\end{cases}\nn
\left (\alpha^{-1}\right)^{(2)} _{\bar s r}&=&0\,.
\eea
The ansatz has four parameters.  The relations, (\ref{keyrelation1}), (\ref{leftinverse}), and (\ref{keyrelation2}) will be enough for us to fix these parameters.  Other relations can be used as additional checks.

Let us begin with the first relation in (\ref{keyrelation1}).  We wish to compute:
\bea
\sum_s\alpha^{(1)}_{qs}\left (\alpha^{-1}\right)^{(1)} _{sq'}&=&{1\over \sqrt{2}}C_{(1)}\delta_{qq'}\nn
&&{}+\!\!\sum_{s\in \mathbb{Z}+\h, s>0} \left [ {1 \over 2\pi i \sqrt{2}}{s+q \over \sqrt{sq}}{1\over s-q}e^{i(s-q)(\sigma_0-\tau_0)}\right ] \left [{C'_{(1)}\over s-q'} \, e^{-i(s-q')(\sigma_0-\tau_0)} \right ] \nn
&=&{1\over \sqrt{2}}C_{(1)}\delta_{qq'}+ {e^{i(q'-q)(\sigma_0-\tau_0)} C'_{(1)}\over 2\pi i \sqrt{2}}\sum_{s\in \mathbb{Z}+\h, s>0} \left [ {s+q \over \sqrt{sq} (s-q)(s-q')}\right ]\nn
\eea
Working in the continuum limit, we would like to approximate the sum by an integral. Note that $q, q'$ are integers, and $s$ in this sum ranges over $s\in \mathbb{Z}+\h$, so the singularities at $s=q, q'$ are automatically regulated. To obtain and compute the required integral, we separate the cases $q = q'$ and $q \neq q'$.

\subsubsection{The case $q=q'$}
We first consider the case $q=q'$. We have
\be
\sum_s\alpha^{(1)}_{qs}\left (\alpha^{-1}\right)^{(1)} _{sq}={1\over \sqrt{2}}C_{(1)}+ { C'_{(1)}\over 2\pi i \sqrt{2}}\sum_{s\in \mathbb{Z}+\h,s>0} \left [ {s+q \over \sqrt{sq} (s-q)^2}\right ]
\ee
The terms in the sum are sharply peaked around the point $s=q$. Thus to leading order we can approximate the factors that are finite and nonvanishing at $s\approx q$ by their value at $s=q$. We then have:
\be\label{firstapprox}
\sum_{s\in \mathbb{Z}+\h,s>0} \left [ {s+q \over \sqrt{sq} (s-q)^2}\right ]\approx 2\sum_{s\in \mathbb{Z}+\h,s>0}  {1\over  (s-q)^2}
\ee

Now in the sum (\ref{firstapprox}) the index $s$ runs over positive half-integers.  Let us then define the quantity:
\be
n \equiv s - q - \tfrac{1}{2}
\ee
Then $n$ runs from $-q$ to $\infty$.  But $q \sim k \gg 1$, and since the denominator is squared we can extend this range to $-\infty$ with only order $k^{-1}$ corrections.  We then have:
\bea
2\sum_{s\in \mathbb{Z}+\h}  {1\over  (s-q)^2} &=& 2\sum_{n = -q}^{\infty}{1\over \left ( n + \h \right )^2} \nn
&\approx& 2\sum_{n = -\infty}^{\infty}{1\over \left ( n + \h \right )^2} ~=~ 2\pi^2
\eea

Thus we find that at order $k^0$, we have:
\be\label{diagonalresult}
\sum_s\alpha^{(1)}_{qs}\left (\alpha^{-1}\right)^{(1)} _{sq}={1\over \sqrt{2}}C_{(1)}+  C'_{(1)}{\pi \over  i \sqrt{2}}.
\ee

\subsubsection{The case $q\neq q'$}
When $q\neq q'$ we have
\be
\sum_s\alpha^{(1)}_{qs}\left (\alpha^{-1}\right)^{(1)} _{sq'}={1\over \sqrt{2}}C_{(1)}+ { C'_{(1)}\over 2\pi i \sqrt{2}}\sum_{s\in \mathbb{Z}+\h} \left [ {s+q \over \sqrt{sq} (s-q)(s-q')}\right ]
\ee
Let us re-write the summand:
\bea\label{split}
{s+q \over \sqrt{sq} (s-q)(s-q')}&=&{(s-q)+2q \over \sqrt{sq} (s-q)(s-q')}={1 \over \sqrt{sq} (s-q')}+{2q \over \sqrt{sq} (s-q)(s-q')}\nn
&=&{1 \over \sqrt{sq} (s-q')}+{2\sqrt{q}\over (q-q')}{1\over \sqrt{s} (s-q)}-{2\sqrt{q}\over (q-q')}{1\over \sqrt{s} (s-q')}\nn
\label{total}
\eea
Each of these terms separately vanishes at order $k^0$ when we sum over $s$.  We will treat the first term in detail below, and the other two may be treated similarly. 

The first term in (\ref{split}) gives the sum
\be
S=\sum_{s\in \mathbb{Z}+\h}{1 \over \sqrt{sq} (s-q')}
\label{sum1}
\ee
The singularity near  $s=0$ is integrable, since $\int {1\over \sqrt{s}}ds \r \sqrt{s}$, so we will not worry about the vicinity of $s=0$. Near $s=q'$, we isolate a range of $s$ around $q'$:
\be
|s-q'|<L
\ee
where $L$ is a momentum scale which is smaller than our large scale $k$, but still large compared to the spacings between the available mode numbers, which are  $\D q= {1 \over M}$, $\D r = {1\over N}$, $\D s = {1\over M+N}$:
\be
L = \e k, \quad \e \ll 1, \quad L\gg \D q, \D r, \D s \,,
\label{range}
\ee
where $\e$ is a fixed small parameter which does not scale with $k$.  In this section, since $M=N=1$ we have $\D q = \D r = 1$, and so the third condition in (\ref{range}) is simply $L \gg 1$.

Inside the above range we will be able to approximate the summand, which in turn allows us to approximate the sum in that region.  Outside of the range we will replace the sum by an integral, since the spacing of points in the sum is small compared to all the remaining scales in the problem.

First, consider the sum arising from $s$ in the range (\ref{range}). In this range we can approximate the other factors in the summand by expanding $s$ around $q'$:
\bea
{1\over \sqrt{sq}} &=& {1\over \sqrt{qq'}}-{1\over 2\sqrt{q q'^3}} (s-q')+\ldots 
\eea
The leading term gives a vanishing contribution, since the summand is antisymmetric while the range of summation is symmetric:
\be
{1\over \sqrt{qq'}}\sum_{s=q'-L+\h}^{q'+L-\h} {1\over (s-q')}=0
\ee
The next subleading term gives a contribution
\be
-{1\over 2\sqrt{q q'^3}} (2L) ~=~ -{1\over \sqrt{qq'}} {L\over q'} ~\sim~ -{1\over \sqrt{qq'}}\epsilon
\label{subleading}
\ee
Recall that we consider $q$ and $q'$ to be of the same order $k$.  Recall also that $\e \ll 1$, which does not scale with $k$.  We then see that (\ref{subleading}) is order ${k^{-1}}$, and can be ignored.

Proceeding in a similar way, we find that the other two terms in (\ref{total}) also give a vanishing contribution at order $k^0$. This leaves us with just with the integral term: 
\be
I={\cal P} \ints_0^\infty ds  {1 \over \sqrt{sq} (s-q')}
\label{integral}
\ee
where the symbol ${\cal P}$ denotes the principal value. This integral runs over the range $s\in [0, \infty)$. We wish to convert this into an integral that runs over the entire real line, so that we may use contour methods. We thus write
\be
s=\tilde s^2, ~~~ ds=2\tilde s d\tilde s
\ee
getting
\be
I={\cal P} \ints_0^\infty d\tilde s   {2 \over \sqrt{q} (\tilde s^2 -q')}
\ee
Since the integrand is symmetric in $\tilde s$, we can write it as
\be
I={\cal P} \ints_{-\infty}^\infty d\tilde s   {1 \over \sqrt{q} (\tilde s^2 -q')}
\label{integralp}
\ee
The integral converges at infinity, so we may close the contour with a semicircle in the upper half plane. We have singularities at $\tilde s=\pm \sqrt{q'}$, which are to be handled as principal values; thus we have to take half the residues at these points. We observe that these residues are equal and opposite:
\bea
\tilde s=\sqrt{q}: ~~Residue&=&{1\over \sqrt{qq'}}\nn
\tilde s=-\sqrt{q}: ~~Residue&=&-{1\over \sqrt{qq'}}
\eea
Thus we find that
\be
I=0
\ee

Putting all of this together, we find that to order $k^0$:
\be
\sum_s\alpha^{(1)}_{qs}\left (\alpha^{-1}\right)^{(1)} _{sq'}=0, \qquad \quad q\ne q'.
\ee
Combined with (\ref{diagonalresult}), we have:
\be\label{qqresult}
\sum_s\alpha^{(1)}_{qs}\left (\alpha^{-1}\right)^{(1)} _{sq'}=\left ({1\over \sqrt{2}}C_{(1)}+  C'_{(1)}{\pi \over  i \sqrt{2}} \right ) \d_{qq'}.
\ee

\subsection{Solving for $C$ and $C'$}
In order for $\a^{-1}$ to be the correct inverse, we must satisfy (\ref{keyrelation1}) and (\ref{keyrelation2}):
\be
\sum_{s}\alpha^{(1)} _{qs}\left (\alpha^{-1}\right)^{(1)} _{sq'} =  \d_{qq'}, \qquad \sum_{s}\alpha^{(2)} _{rs}\left (\alpha^{-1}\right)^{(2)} _{sr'} =  \d_{rr'} \nonumber
\ee
\be
\sum_{s}\alpha^{(1)} _{qs}\left (\alpha^{-1}\right)^{(2)} _{sr} =  \sum_{s}\alpha^{(2)} _{rs}\left (\alpha^{-1}\right)^{(1)} _{sq} =  0 
\ee
We have explicitly calculated the sum in the first relation.  The same methods can be used for each of the other sums in (\ref{keyrelation1}) and (\ref{keyrelation2}).  This yields:
\bea
\sum_s\alpha^{(2)}_{rs}\left (\alpha^{-1}\right)^{(2)}_{sr'}&=&\left [{1\over \sqrt{2}}C_{(2)}-  C'_{(2)}{\pi \over  i \sqrt{2}}\right ]\delta _{rr'}\nn
\sum_s\alpha^{(1)}_{qs}\left (\alpha^{-1}\right)^{(2)}_{sr}&=&\left [{1\over \sqrt{2}}C_{(2)}+  C'_{(2)}{\pi \over  i \sqrt{2}}\right ]\delta _{qr}\nn
\sum_s\alpha^{(2)}_{rs}\left (\alpha^{-1}\right)^{(1)}_{sq}&=&\left [{1\over \sqrt{2}}C_{(1)}-  C'_{(1)}{\pi \over  i \sqrt{2}}\right ]\delta _{rq}
\eea
This equation along with (\ref{qqresult}) gives us a system of four equations with four unknowns.  Solving this system yields:
\bea
C_{(1)} & = & C_{(2)} = {1\over \sqrt{2}} \nn
C_{(1)}' & = & -C_{(2)}' = -{1\over \pi i\sqrt{2}}
\eea
which we plug back into our ansatz to find:
\be
\left (\alpha^{-1}\right)^{(1)} _{sq}= \begin{cases}
{1\over \sqrt{2}}\delta_{sq} & s \in \mathbb{Z} \\
-{1\over \pi i\sqrt{2}}{1\over s-q} \, e^{-i(s-q)(\sigma_0-\tau_0)} & s \in \mathbb{Z}+\h
\end{cases}
\ee
and
\be
\left (\alpha^{-1}\right)^{(2)} _{sr}= \begin{cases}
{1\over \sqrt{2}}\delta_{sr} & s \in \mathbb{Z} \\
{1\over \pi i \sqrt{2}}{1\over s-r} \, e^{-i(s-r)(\sigma_0-\tau_0)} & s \in \mathbb{Z}+\h \,.
\end{cases}
\ee

In a similar manner we can also check the other required relations.  In particular, we note that $\alpha^{-1}$ also functions as a left-inverse of $\alpha$, up to order $k^0$.

\section{Calculating $\g^B$}\label{1,1gamma}
For the matrix $\g^B = \a^{-1}\beta$, there are two contributions to the holomorphic quantity $\g^B_{ss'}$:
\be
\g^B_{ss'} = \sum_{q}\left (\alpha^{-1}\right)^{(1)} _{sq}\beta^{(1)}_{qs'} + \sum_{r}\left (\alpha^{-1}\right)^{(2)} _{sr} \beta^{(2)}_{rs'}
\label{gamma1}
\ee
We will define these two contributions as $\g^{B(1)}_{ss'}$ and $\g^{B(2)}_{ss'}$ respectively.  Because of the piecewise-defined nature of $\a$ and $\beta$, we will have to separate the cases of $s$ and $s'$ being integers or half-integers.  Fortunately, $\beta = 0$ when $s'$ is an integer, so we can only have nonzero $\g^B$ when $s'$ is half-integer.  With this constraint, we are left with the two cases of $s$ integer and $s$ half-integer.

When $s$ is an integer, $\a^{-1}$ is non-zero only when $q=s$.  We then have:
\bea
\g^B_{ss'} &=& \left (\alpha^{-1}\right)^{(1)} _{ss}\beta^{(1)}_{ss'} + \left (\alpha^{-1}\right)^{(2)} _{ss}\beta^{(2)}_{ss'} \nn
&=& {1\over \sqrt{2}}\beta^{(1)}_{ss'} + {1\over \sqrt{2}}\beta^{(2)}_{ss'}
\eea
But we know from (\ref{beta1}) and (\ref{beta2}) that $\beta^{(2)}_{ss'} = -\beta^{(1)}_{ss'}$, and thus the two contributions cancel.  We thus find that $\g^B_{ss'}$ is only nonzero when both $s$ and $s'$ are half-integer.

When $s$ and $s'$ are both half-integers, the $\d_{qs}$ and $\d_{rs}$ parts of $\a^{-1}$ do not contribute.  We thus have:
\bea
\g^B_{ss'} & = & {1 \over 4\pi^2}e^{-i(s+s')(\s_0-\t_0)}\bigg{(}\sum_{q=1}^{\infty}{1\over s-q}{s'-q \over \sqrt{s'q}(s'+q)} + \sum_{r=1}^{\infty}{1\over s-r}{s'-r \over \sqrt{s'r}(s'+r)}\bigg{)}\quad
\eea
where the fact that there is a sign difference in both $\a^{-1}$ and $\b$ between the $(1)$ and $(2)$ sectors for half-integer $s$ causes the two sums to enter with the same sign.  Now $q$ and $r$ are both summed over the positive integers, so our two sums are actually identical.  We thus have:
\bea
\g^B_{ss'} &=& {1 \over 2\pi^2}e^{-i(s+s')(\s_0-\t_0)}\sum_{q}{1\over s-q}{s'-q \over \sqrt{s'q}(s'+q)}
\eea

Again we take a symmetric box around the pole and treat it carefully, while evaluating the rest of the sum as a principal value integral.  With $L = \e k$ as before, we have:
\bea
\g^B_{ss'} &=& {1 \over 2\pi^2}e^{-i(s+s')(\s_0-\t_0)}\bigg{(}\sum_{q = s-L+\h}^{s+L-\h}{1\over s-q}{s'-q \over \sqrt{s'q}(s'+q)} \nn
&&{}+ {\cal P} \ints_{0}^{\infty}{dq\over s-q}{s'-q \over \sqrt{s'q}(s'+q)} \bigg{)}
\eea
In the summed over region, $q \approx s$, and so the sum is approximately odd around the pole.  It therefore gives a subleading contribution.  Turning to the integral, we perform the substitution $\tilde{q}^2 = q$.  This yields an even integrand, allowing us to extend the integral to the entire real axis.  We then have:
\bea
\g^B_{ss'} &=&{1 \over 2\pi^2}e^{-i(s+s')(\s_0-\t_0)}{\cal P} \ints_{-\infty}^{\infty}{d\tilde{q}\over s-\tilde{q}^2}{s'-\tilde{q}^2 \over \sqrt{s'}(s'+\tilde{q}^2)}
\eea
where the poles on the real axis at $\pm \sqrt{s}$ are treated with the principal value prescription.  The residues from these poles exactly cancel.  However, there are also poles at $\pm i \sqrt{s'}$.  Closing in the upper-half plane, we only enclose the $+i\sqrt{s'}$ pole.  We thus find that when $M=N=1$ and $s$ and $s'$ are both half-integer, the matrix $\g^B$ is:
\bea
\g^B_{ss'}&=&{1 \over \pi(s+s')}e^{-i(s+s')(\s_0-\t_0)}
\eea
This agrees with the continuum limit result of the corresponding quantity computed in \cite{acm2}.\footnote{In \cite{acm2}, the bosonic excitations were described by the operators $\a_m$, which have the commutation relation $\left [ \a_{m},\a_n \right ] = m\d_{m+n,0}$.  Our bosonic modes are instead canonically normalized.  This means that there is a factor of the square root of the mode number between our modes and the modes used in \cite{acm2}.  That is, $a_q = \sqrt{|q|}\;\a_q$.  Since $\g^B$ accompanies a pair of bosonic modes, our gamma has an extra factor of $\sqrt{ss'}$ relative to the quantity computed in \cite{acm2}.}

\newpage

\section{The $\a$, $\b$, and $\a^{-1}$ Matrices for General $M$ and $N$}\label{generalMN}
We now generalize the results of the previous two sections to the case of arbitrary $M$ and $N$.  Before the twist we have two component strings.  The first has winding number $M$, while the second has winding number $N$.  We have a field $X^{(1)}$ with period $2\pi M$ on the first component string, and a field $X^{(2)}$ with period $2\pi N$ on the second component string.  The mode functions for these fields are:
\begin{align}
f^{(1)}_{q} & ~=~ {1\over \sqrt{2 \pi M}}{1\over \sqrt{2q}}e^{i q (\s - \t)}  & 0 &\leq \s < 2\pi M\\
f^{(1)}_{\bar q} & ~=~ {1\over \sqrt{2 \pi M}}{1\over \sqrt{2 \bar q}}e^{-i q (\s + \t)}  & 0 &\leq \s < 2\pi M\\
f^{(2)}_{r} & ~=~ {1\over \sqrt{2 \pi N}}{1\over \sqrt{2r}}e^{i r (\s - \t)} \qquad & 0 &\leq \s < 2\pi N\\
f^{(2)}_{\bar r} & ~=~ {1\over \sqrt{2 \pi N}}{1\over \sqrt{2 \bar r}}e^{-i \bar r (\s + \t)} \qquad  & 0 &\le \s < 2\pi N
\end{align}
where $q$ and $\bar q$ are multiples of ${1 \over M}$, and $r$ and $\bar r$ are multiples of ${1 \over N}$.  After the twist we have a single component string with winding number $M+N$, and a field $X$ with period $2\pi (M+N)$.  The mode functions for this field are:
\begin{align}
f_{s} & ~=~ {1\over \sqrt{2 \pi (M+N)}}{1\over \sqrt{2s}}e^{i s (\s - \t)} \qquad & 0 &\leq \s < 2\pi (M+N)\\
f_{\bar s} & ~=~ {1\over \sqrt{2 \pi M+N)}}{1\over \sqrt{2 \bar s}}e^{-i \bar s (\s + \t)} \qquad & 0 &\leq \s < 2\pi (M+N)	
\end{align}
where $s$ and $\bar s$ are multiples of ${1 \over M+N}$.

We again match the fields at the time of the twist insertion.  Starting at $\s = 0$ on the cylinder, we set the field $X$ on the final component string to be equal to the field $X^{(1)}$ from the first component string.  As before, we move around the cylinder until we reach $\s = \s_0$.  Here the twist moves us to the second copy of the initial component strings, and we have $X = X^{(2)}$.  Now we continue moving around the cylinder until we return to the twist insertion \emph{on the second component string}, which requires us to traverse the component string's full period $2\pi N$.  Thus we return to the twist insertion at $\s = \s_0 + 2\pi N$.  Here the twist moves us back to first copy of the initial component strings, so we again have $X = X^{(1)}$.  This gives the relations:
\bea
X\big{(}0 \leq \s < \s_0, \t_0\big{)} & = & X^{(1)}(\s,\t_0) \nn
X\big{(}\s_0 \leq \s < 2\pi N + \s_0, \t_0\big{)} & = & X^{(2)}(\s,\t_0) \nn
X\big{(}2\pi N + \s_0 \leq \s < 2\pi(M+N), \t_0\big{)} & = & X^{(1)}(\s - 2\pi N,\t_0) \nn
\eea
Here the field $X^{(1)}$ does not in general have the same period as the field $X^{(2)}$.  Thus the shift in the coordinate of this field in the third relation is non-trivial, and we must account for it explicitly.  With these relations, the inner product (\ref{inner}) takes the form:

\bea\label{innerproductmn}
(f_s,f^{(1)}_q)&=&-i\ints_{0}^{\s_0} d\sigma \left (f_s\,  \p_\tau f^{(1)*}_q - \p_\tau f_s \, f^{(1)*}_q\right ) -i\ints_{2\pi N + \s_0}^{2\pi(M+N)} d\sigma \left (f_s\,  \p_\tau f^{(1)*}_q - \p_\tau f_s \, f^{(1)*}_q\right ) \nn
(f_s,f^{(2)}_r)&=&-i\ints_{\s_0}^{2\pi N + \s_0} d\sigma \left (f_s\,  \p_\tau f^{(2)*}_r - \p_\tau f_s \, f^{(2)*}_r\right )
\eea
and similarly for the anti-holomorphic mode functions.

Before proceeding to the calculation of the Bogoliubov coefficients, it is helpful to set up some notation.  In the $M=N=1$ case, we saw that when indices take particular values, such as the cases $q=s$, $r=s$, and so on, separate treatment was typically required.  This feature will also be present in the general $M$ and $N$ case, so we need to identify when these equalities occur.  In general, $M$ and $N$ may share common factors.  We therefore write:
\be
Y = \gcd(M,N), ~~~ m = {M\over Y}, ~~~ n = {N \over Y}
\ee
where $m$ and $n$ are now coprime.  With this we introduce the notation:
\bea
q & = & q_0 + {q_1 \over Y} + {q_2 \over M} \\
r & = & r_0 + {r_1 \over Y} + {r_2 \over N} \\
s & = & s_0 + {s_1 \over Y} + {s_2 \over M+N}
\eea
where all quantities on the right side are integers, and we have the constraints:
\bea
q_0,r_0,s_0 &\geq& 0 \nn
q_1,r_1,s_1&\in&\{ 0,1,\ldots,Y-1\} \nn
q_2&\in&\{ 0,1,\ldots,m-1\} \nn
r_2&\in&\{ 0,1,\ldots,n-1\} \nn
s_2&\in&\{ 0,1,\ldots,m+n-1\}
\eea
In this way, every allowed value has a unique representation.  It is then apparent that in order to have have say $q=s$, we require $q_0 = s_0$, $q_1 = s_1$, and $q_2 = s_2 = 0$.  We will also impose the constraint $q,r,s > 0$, since we are not working with zero modes.  We use the same notation for the anti-holomorphic indices.

We further introduce the commonly occurring factor\footnote{We could have just as easily defined the factor in terms of $N$ using the relation $e^{2\pi i M s} = e^{-2\pi i N s}$.}
\be
\mu_s = 1 - e^{2\pi i M s} = 1 - e^{2\pi i {M s_2 \over M+N}} \label{mu}
\ee
This factor is zero when $s_2 = 0$, and thus it vanishes when equalities like $q=s$ are satisfied.  This allows us to handle equalities like $q=s$ by pulling the $s_2=0$ terms out of our sums.  In the case of $M=N=1$, this factor vanished for integer $s$ and evaluated to $2$ for half-integer $s$.

With this new notation, one can determine all of the pieces of the $\a$ and $\b$ matrices.  There are many such pieces, and only those most relevant to our calculation are shown here.  We also record here the most relevant pieces of $\a^{-1}$ in the continuum limit. A full compilation of all the pieces is given in Appendix \ref{fullresults}.
\bea\label{alpha1}
\a^{(1)}_{qs} & = & \left ( f_s , f^{(1)}_q \right ) ~=~ 
\begin{cases}
\sqrt{M\over M+N}\d_{qs} & \qquad s_2 = 0 \\
{1 \over 4\pi i \sqrt{M(M+N)}}{s+q \over \sqrt{sq}}{1\over s-q}\mu_{-s}e^{i(s-q)(\s_0-\t_0)} & \qquad s_2 \neq 0
\end{cases} \nn
\a^{(1)}_{q \bar s} & = &\left ( f_{\bar s} , f^{(1)}_q \right ) ~=~ -{1 \over 4\pi i \sqrt{M(M+N)}}{1\over\sqrt{\bar s q}} \mu_{\bar s}e^{-i(\bar s+ q)\s_0} e^{-i(\bar s - q) \t_0}
\eea

\bea
\b^{(1)}_{qs} &=& \left ( f^*_s , f^{(1)}_q \right ) ~=~ {1 \over 4\pi i \sqrt{M(M+N)}}{s-q \over \sqrt{sq}}{1\over s+q}\mu_se^{-i(s+q)(\s_0-\t_0)} \nn
\b^{(1)}_{q \bar s} & = &\left ( f^*_{\bar s} , f^{(1)}_q \right ) ~=~ -{1 \over 4\pi i \sqrt{M(M+N)}}{1\over\sqrt{\bar s q}} \mu_{-\bar s}e^{i(\bar s- q)\s_0}e^{i(\bar s +  q)\t_0} \label{betamn}
\eea

\bea
(\a^{-1})^{(1)}_{sq} & = & f^{(1)}_{qs} ~=~ 
\begin{cases}
\sqrt{M\over M+N}\d_{sq} & \qquad s_2 = 0 \\
-{1\over 2\pi i \sqrt{M(M+N)}}{1\over s-q}\m_{s}e^{-i(s-q)(\s_0-\t_0)} & \qquad s_2 \neq 0
\end{cases} \nn
(\a^{-1})^{(1)}_{\bar s q} & = & f^{(1)}_{q \bar s} ~=~ 0 \label{alphainvmn}
\eea

\bea
\a^{(2)}_{rs} & = &\left ( f_s , f^{(2)}_r \right ) ~=~ 
\begin{cases}
\sqrt{N\over M+N}\d_{rs} & \qquad s_2 = 0 \\
-{1 \over 4\pi i \sqrt{N(M+N)}}{s+r \over \sqrt{sr}}{1\over s-r}\mu_{-s}e^{i(s-r)(\s_0-\t_0)} & \qquad s_2 \neq 0
\end{cases} \nn
\a^{(2)}_{r \bar s} & = &\left ( f_{\bar s} , f^{(2)}_q \right ) ~=~ {1 \over 4\pi i \sqrt{N(M+N)}}{1\over\sqrt{\bar s r}} \mu_{\bar s}e^{-i(\bar s+ r)\s_0} e^{-i(\bar s - r) \t_0}
\eea

\bea
\b^{(2)}_{rs} &=& \left ( f^*_s , f^{(2)}_r \right ) ~=~ -{1 \over 4\pi i \sqrt{N(M+N)}}{s-r \over \sqrt{sr}}{1\over s+r}\mu_se^{-i(s+r)(\s_0-\t_0)} \nn
\b^{(2)}_{r \bar s} & = &\left ( f^*_{\bar s} , f^{(2)}_r \right ) ~=~ {1 \over 4\pi i \sqrt{N(M+N)}}{1\over\sqrt{\bar s r}} \mu_{-\bar s}e^{i(\bar s- r)\s_0}e^{i(\bar s +  r)\t_0}
\eea

\bea
(\a^{-1})^{(2)}_{sr} & = & f^{(2)}_{rs} ~=~
\begin{cases}
\sqrt{N\over M+N}\d_{sr} & \qquad s_2 = 0 \\
{1\over 2\pi i \sqrt{N(M+N)}}{1\over s-r}\m_{s}e^{-i(s-r)(\s_0-\t_0)} & \qquad s_2 \neq 0
\end{cases}\nn
(\a^{-1})^{(2)}_{\bar s r} & = & f^{(2)}_{r \bar s} ~=~ 0
\eea
We have explicitly verified, to leading order in our approximation, that
\bea
\a\a^{-1} &=& \a^{-1}\a ~=~ \one \,.
\eea
The calculation is somewhat lengthy, and is presented in Appendix \ref{InverseCheck}. We have also explicitly verified that the quantity $f^B$ defined in \eq{eq:fbdef} satisfies:
\bea
f^B &\equiv& \a^* - \b^* \g^B ~=~ \left (\a^{-1}\right )^T \,.
\eea

Let us make a comment on the form of the above expressions for the various Bogoliubov coefficients.  Consider, for example, $\a^{(1)}_{qs}$ for $s_2 \neq 0$, as given in (\ref{alpha1}).  The factor $\mu_{-s}$ causes this expression to vanish whenever $s_2 = 0$ except when $s = q$, in which case the factor of $s-q$ in the denominator also vanishes.  In taking the limit of this expression as $s \to q$, one obtains the result $\sqrt{\tfrac{M}{M+N}}$, which matches the value calculated when $q=s$.  In other words, the expression for $\a^{(1)}_{qs}$ when $s_2 = 0$ can be obtained from the expression when $s_2 \neq 0$ by taking the limit as $s_2 \to 0$.  This observation holds for all the Bogoliubov coefficients studied in this section.

\section{Calculating $\g^B$ for general $M,N$}\label{gammaMN}

We now compute the quantity $\g^B$ for the case of general $M$ and $N$.   We will continue to work only to leading order in the small parameters $\tfrac{1}{(M+N)k}$ and $\e$. As in the case of $M=N=1$, the matrix $\g^B$ receives contributions from the $q$ modes on the first component string and the $r$ modes on the second component string,
\be
\g^B_{ss'} ~=~ \left [\a^{-1}\b\right ]_{ss'} = \sum_{q}(\a^{-1})^{(1)}_{sq}\b^{(1)}_{qs'} + \sum_{r}(\a^{-1})^{(2)}_{sr}\b^{(2)}_{rs'} ~\equiv~ \g^{B(1)}_{ss'} + \g^{B(2)}_{ss'}
\label{gamma2}
\ee
Let us compute these two contributions separately.  Both contributions clearly vanish when $s_2'=0$, as the components of the $\b$ matrix vanish in this case.  But since $\a^{-1}$ is described piecewise, we must handle the cases of $s_2 = 0$ and $s_2 \neq 0$ separately.

\subsection{The $s_2 = 0$ Case}
When $s'_2 = 0$, the entire $\b$ vanishes, and thus $\g^B$ is zero.  $\g^B$ should be symmetric, so let us check that $\g^B_{ss'}$ is zero when $s_2 = 0$.  When $s_2 = 0$, $\a^{-1}$ vanishes except when $q = s$.  Thus only one term from each sum contributes.  We then have:
\bea
\g^{B(1)}_{ss'} & = & \sqrt{M \over M+N}{1\over 4\pi i \sqrt{M(M+N)}}{s'-s \over \sqrt{ss'}}{1\over s'+s}\mu_{s'}e^{-i(s+s')(\s_0 - \t_0)} \nn
&=& {1\over 4\pi i (M+N)}{s' - s \over \sqrt{ss'}}{1\over s' + s}\mu_{s'}e^{-i(s+s')(\s_0 - \t_0)}
\eea
while
\bea
\g^{B(2)}_{ss'} & = & \sqrt{N \over M+N}{-1\over 4\pi i \sqrt{N(M+N)}}{s'-s \over \sqrt{ss'}}{1\over s'+s}\mu_{s'}e^{-i(s+s')(\s_0 - \t_0)} \nn
&=& -{1\over 4\pi i (M+N)}{s' - s \over \sqrt{ss'}}{1\over s' + s}\mu_{s'}e^{-i(s+s')(\s_0 - \t_0)} \nn
&=& - \g^{B(1)}_{ss'}
\eea
Which in turn tells us that:
\be
\g^B_{ss'} = 0 \qquad s_2 \text{ or }s'_2 = 0.
\ee

\subsection{$\g^{B(1)}$ for $s_2 \neq 0$}
We now consider the more interesting case of $s_2 \neq 0$.  We first compute $\g^{B(1)}_{ss'}$.  Plugging in $\a^{-1}$ and $\b$ for $s_2 \neq 0$ from (\ref{alphainvmn}) and (\ref{betamn}), we find:
\bea
\g^{B(1)}_{ss'} & = & {1\over M}C_{s,s'}\sum_q {s'-q \over \sqrt{s'q}}{1\over s'+q}{1\over s-q}
\eea
where $q$ ranges over the positive multiples of ${1\over M}$, and where
\bea\label{css}
C_{s,s'} & \equiv & {1\over 8\pi^2(M+N)}e^{-i(s+s')(\s_0-\t_0)}\mu_s\mu_{s'}
\eea

The above expression for $\g^{B(1)}_{ss'}$ would be divergent if $q$ could attain the value $q=s$, due to the pole in the summand.  However, since we have $s_2 \neq 0$, $q$ never obtains this value exactly.   For $M=N=1$, we saw that the order $k^0$ contribution from this region vanished due to symmetry.  However, in the present case the values of $q$ can in general be very asymmetric about the pole when $q \approx s$, since $s-q$ can have a non-integer part which takes a different range of values for $q>s$ than for $q<s$.  Thus the contribution from this region does not vanish by symmetry as it did in the $M=N=1$ case.

Since $s$ and $s'$ are of order $k$ which is large compared to $\tfrac{1}{M+N}$, let us again cut out a region of $\pm L$ around this pole, where $L = \e k$ as before.  To specify the limits of the sum, let us identify $\tilde q$ as the value of $q$ which minimizes the distance $|s-q|$.  We will then write:
\be
\d = s - \tilde q
\ee
We then take the bounds of the sum to be symmetric around $\tilde q$:
\bea
q_{min} &=& \tilde q - L ~=~ s - \d - L \nn
q_{max} &=& \tilde q + L ~=~ s - \d + L
\eea

Recall from \eq{range} that we have $L \gg \D q$.  Thus we are approaching the pole in an approximately symmetric fashion when outside the interval $[s-L,s+L]$.  We can therefore use the principal value when outside of this region.  We will later show that the principal value integral gains no contribution from the pole at $q = s$.  We can therefore write:
\bea
\g^{B(1)}_{ss'} &\approx& {1\over M}C_{s,s'}\bigg{(} {\cal P}\ints_0^{\infty}\Big{(}{s'-q \over \sqrt{s'q}}{1\over s'+q}{1\over s-q} \Big{)}M\diff q + \sum_{q=s-L}^{s+L}{s'-q \over \sqrt{s'q}}{1\over s'+q}{1\over s-q}\bigg{)} \nn
&\equiv& I + S
\eea
where $I$ is the contribution from the integral and $S$ is the contribution from the sum.  The factor of $M$ in the principal value integral comes from the fact that $q$ is being summed over integer multiples of $1/M$.

\subsubsection{The Principal Value Integral}
To calculate the principal value integral, we perform the change of variables $\tilde{q}^2 = q$, giving $\sqrt{q} = \tilde q,$ and $\diff q = 2\tilde{q} \diff \tilde{q}$.  We then obtain:
\bea
I & = & 2C_{s,s'}{\cal P}\ints_{0}^{\infty} {s' - \tilde{q}^2 \over \sqrt{s'}}{1\over s' + \tilde{q}^2}\Big{(}{1\over s - \tilde{q}^2}\Big{)}\diff \tilde{q}
\eea
The integrand is even, so we extend the range to $-\infty$.  The integrand also behaves as $\tilde{q}^{-2}$ for large $\tilde q$, so we can close with a contour in the upper half-plane which integrates to zero.  We then have an integral over a contour $C_+$, which completely encloses the upper half-plane:
\bea
I & = & {C_{s,s'}\over \sqrt{s'}}{\cal P}\int\limits_{C_+}{s' - \tilde{q}^2 \over s' + \tilde{q}^2}\Big{(}{1 \over s - \tilde{q}^2}\Big{)} \diff \tilde{q}
\eea

The poles in the upper half-plane are found at $\tilde{q} = i\sqrt{s'},$ $\tilde{q}= \sqrt{s} + i\e,$ and $\tilde{q} = -\sqrt{s} + i\e$.  Residues for the last two come with a factor of $\h$ due to the principal value prescription.  We thus have:
\bea\label{pvintegral}
I &=& {C_{s,s'}\over \sqrt{s'}}2\pi i \bigg{(}{2s' \over 2i\sqrt{s'}}{1\over s+s'} + \h\Big{(} {s' - s \over s' + s}{1\over 2\sqrt{s}} +{s' - s \over s' + s}{1\over (-2)\sqrt{s}}\Big{)}\bigg{)} \nn
&=& {2\pi C_{s,s'} \over s+s'}
\eea
And we see that the pole at $q = s$ indeed gives no contribution to the principal value integral.

Plugging in the value for $C_{s,s'}$, we find that the integral $I$ gives a contribution of order $\tfrac{1}{(M+N)k}$ to the quantity $\g^{B(1)}_{ss'}$. This will turn out to be the leading-order answer.  Thus when evaluating the contribution from the sum, we will drop terms that are subleading relative to this contribution.

\subsubsection{The sum over the near-pole region}
Now we turn to the sum over the near-pole region.  Since $s$ and $s'$ are much larger than $L$, the index $q$ remains near $s$ over the entire region.  Thus for each factor except the pole, we can expand around $q = s$.  We then have:
\bea\label{sumexpansion}
S & \approx & {1\over M}C_{s,s'} \sum_{q=\tilde q-L}^{\tilde q+L}\left ({s'-s \over \sqrt{s's}(s'+s)}{1\over s - q} + \mathcal{O}\left ( \tfrac{1}{k^2} \right ) + (s-q)\mathcal{O}\left ( \tfrac{1}{k^3} \right ) + \ldots \right )
\eea
where $C_{s,s'}$ was defined in (\ref{css}) and is of order $\tfrac{1}{M+N}$. Only the coefficient of the zeroth order term in the expansion has been written explicitly. The coefficients of the remaining terms contain further powers of both $\tfrac{1}{k}$ and $s-q$.  These elements are suppressed by successive powers of $k$, and after summing over a range of order $L=\e k$, these terms will be suppressed by successive powers of $\e$ relative to the leading term.  Let us now demonstrate this.

All of our terms have the form:
\be
{\mathcal{K}\over (M+N)}\sum_{s-q = \d -L}^{\d + L} {(s-q)^m \over k^{m+2}}
\ee
where $m$ takes the values $-1, 0, 1, \ldots$, and $\mathcal{K}$ is some overall factor that is not necessarily suppressed in any way.  With $L = \e k$, the $m = -1$ term, which is the first term in (\ref{sumexpansion}), will behave as $\tfrac{1}{(M+N)k}\log(\e k)$.  Thus at this level it is not obviously suppressed relative to the contribution (\ref{pvintegral}) from the principal value integral.  We will return to this term momentarily.  But first, let us consider the $m\geq 0$ terms, which take the form:
\bea
{1\over (M+N)}\sum_{s-q = \d - L}^{\d + L} {(s-q)^m \over k^{m+2}} &\sim & {1\over (M+N)k}\left ({L \over k} \right )^{m+1} \nn
&\sim & {1\over (M+N)k}\e^{m+1}
\eea
These terms are suppressed by powers of $\e$ relative to the contribution (\ref{pvintegral}) from the principal value integral, so we will discard them and work only with the $m=0$ term.  We then have:
\bea\label{Sapprox}
S & \approx & {1\over M}C_{s,s'} {s'-s \over \sqrt{s's}(s'+s)}\sum_{q=\tilde q - L}^{\tilde q+L}{1\over s - q}~\equiv~ D_{s,s'}\sum_{q = \tilde q - L}^{\tilde q+L}{1\over M(s - q)}
\eea
where
\bea
D_{s,s'} & \equiv &C_{s,s'} {s'-s \over \sqrt{s's}(s'+s)}
\eea

To treat this term carefully, we re-write $M(s-q)$ using our standard decomposition of $s$ and $q$:
\bea
M(s-q) & = & M\Big{(}s_0 + {s_1\over Y} + {s_2 \over M+N} - q_0 - {q_1 \over Y} - {q_2 \over M}\Big{)} \nn
&=& Ms_0 + ms_1 - Mq_0 - mq_1 - q_2 + {Ms_2 \over M+N} \nn
&\equiv& n_0 + {Ms_2 \over M+N}
\eea
where $n_0$ is an integer.  Now the bounds of the sum are approximately $n_0 = \pm ML$.  Thus to leading order we have:
\be \label{gammasum}
S \approx D_{s,s'} \sum_{n_0=-M L}^{n_0=M L}{1\over n_0 + {Ms_2 \over M+N}}
\ee

We now pull out the $n_0 = 0$ term, and then pair off each positive value of $n_0$ with the corresponding negative value $-n_0$.  We then have:
\bea
S &\approx& D_{s,s'}\left ( {M+N \over Ms_2} + \sum_{n_0 = 1}^{ML}\left ({1\over n_0 + {Ms_2 \over M+N}} - {1\over n_0 - {Ms_2 \over M+N}}\right ) \right )\nn
&=&D_{s,s'}\bigg{(} {M+N \over Ms_2} + \sum_{n_0 = 1}^{ML}{-2{Ms_2 \over M+N} \over n_0^2 - \big{(}{Ms_2 \over M+N}\big{)}^2}\bigg{)}
\eea
Now the remaining summand falls off as $n_0^{-2}$, which is $\mathcal{O}\left ( {1 \over M^2 L^2}\right )$.  And since $L \gg \D q = \tfrac{1}{M}$, extending the sum to infinity brings corrections of order ${1 \over ML} \ll 1$.  Thus to leading order, we have:
\be
S \approx D_{s,s'}\bigg{(} {M+N \over Ms_2} + \sum_{n_0 = 1}^{\infty}{-2{Ms_2 \over M+N} \over n_0^2 - \big{(}{Ms_2 \over M+N}\big{)}^2}\bigg{)} \label{Ssum}
\ee
We then use the identity:
\be
\sum_{n_0=1}^{\infty}{-2a \over n_0^2 - a^2} = \pi \cot(\pi a) - {1\over a}
\ee
with
\be
a={Ms_2 \over M+N}
\ee
The $-1/a$ term exactly cancels the term that came from $n_0 = 0$, and we are left with:
\be \label{eq:sans}
S ~\approx~ D_{s,s'}\pi \cot \Big{(}\pi {Ms_2 \over M+N} \Big{)} ~=~ C_{s,s'} {s'-s \over \sqrt{s's}(s'+s)}\pi \cot \Big{(}\pi {Ms_2 \over M+N} \Big{)}
\ee
The prefactor is of the same order as the contribution from the principal value integral (\ref{pvintegral}), and thus this term cannot be dropped.  However, we will soon see that it cancels with the corresponding term in $\g^{B(2)}$.

\subsection{$\g^{B(2)}$ for $s_2 \neq 0$}
Plugging in $\a^{-1}$ and $\b$ for the case $s_2 \neq 0$ yields:
\bea
\g^{B(2)}_{ss'} & = & {1\over N}C_{s,s'}\sum_r {s' - r \over \sqrt{s'r}}{1\over s'+r}{1\over s-r}
\eea
where we have the same $C_{s,s'}$ given in (\ref{css}).  This is easy to see from the relations:
\bea
(\a^{-1})^{(2)}_{rs} & = & -\sqrt{M\over N}\Big{[}(\a^{-1})^{(1)}_{qs}\Big{]}_{q \to r} \\
\b^{(2)}_{rs} & = & -\sqrt{M\over N}\Big{[}\b^{(1)}_{qs}\Big{]}_{q \to r}
\eea
Thus for each value of $r$, the contribution to $\g^{B(2)}_{ss'}$ from that value of $r$ is equal to $M/N$ times the contribution $\g^{B(1)}_{ss'}$ would have had from the same value of $q$, if such terms were found in the sum.  Of course, this does not mean that $\g^{B(2)}_{ss'} = {M\over N}\g^{B(1)}_{ss'}$, because the $r$ sum typically ranges over a completely different domain than the $q$ sum.\footnote{The exception is when $M=N$, in which case we \emph{do} have $\g^{B(2)}_{ss'} = \g^{B(1)}_{ss'}$.  In this case, the GCD $Y$ is equal to $M$ (and $N$), and thus $s_2 = 1$ when nonzero.  Then the correction from the near-pole region vanishes.  This is completely expected because in such a case our sum \emph{is} symmetric about the pole, and thus the principal value works just fine.}

We now handle the sum in exactly the same way as before, by treating a large region around the pole carefully and using the principal value everywhere else.  We will again find that the pole at $r=s$ gives no contribution to the principal value integral, and so we can extend that integral along the entire real axis.  We then have:
\bea
\g^{B(2)}_{ss'} &\approx& {1\over N}C_{s,s'}\bigg{(} {\cal P}\ints_0^{\infty}\Big{(}{s'-r \over \sqrt{s'r}}{1\over s'+r}{1\over s-r} \Big{)}N\diff r + \sum_{r=s-L}^{s+L}{s'-r \over \sqrt{s'r}}{1\over s'+r}{1\over s-r}\bigg{)} \nn
&\equiv& \tilde{I} + \tilde{S}
\eea
where the factor of $N$ in the principal value integral comes from the fact that $r$ is being summed over integer multiples of $1/N$.

\subsubsection{The Principal Value Integral}
After cancelling the factors of $N$, we have:
\be
\tilde{I} = C_{s,s'}{\cal P}\ints_0^{\infty}{s'-r \over \sqrt{s'r}}{1\over s'+r}{1\over s-r}\diff r
\ee
Here $r$ is integrated over the continuous region $[0,\infty)$.  This shift to an integral washes out the difference in domains between the $r$ sum and the $q$ sum, so we see that:
\be
\tilde I = I ={2\pi C_{s,s'} \over s+s'}
\ee
and again the real pole gives zero contribution to the principle value integral.

\subsubsection{The sum over the near-pole region}
Again the fact that $s,s' \gg L$ allows us to apply $r \approx s$ over the near-pole region.  This expansion takes the same functional form as the expansion performed in the sum over $q$ in \eq{sumexpansion}, so we know that only the first term in the expansion gives a leading order contribution to the sum.  We thus have:
\bea
\tilde{S} &\approx& {1\over N}C_{s,s'}{s'-s \over \sqrt{s's}(s'+s)}\sum_{r=s-L}^{s+L}{1\over s-r} ~=~ D_{s,s'} \sum_{r=\tilde r-L}^{\tilde r+L}{1\over N(s-r)}
\eea
where the factor $D_{s,s'}$ is the same as before, and we now use $\tilde r$ to denote the value of $r$ which minimizes the distance $|s-r|$.

Now let us re-write $N(s-r)$ using our standard decomposition of $s$ and $r$:
\bea
N(s-r) & = & N\Big{(}s_0 + {s_1 \over Y} + {s_2 \over M+N} - r_0 - {r_1 \over Y} - {r_2 \over N}\Big{)} \nn
&=& Ns_0 + ns_1 - Nr_0 - nr_1 - r_2 + {Ns_2 \over M+N} \nn
& \equiv & \tilde{n}_0 + {Ns_2 \over M+N}
\eea
where we have grouped all of our integers into $\tilde{n}_0$, which ranges approximately from $-NL$ to $NL$.  Thus to leading order we have:
\be
\tilde{S} = D_{s,s'}\sum_{\tilde{n}_0=-NL}^{NL}{1\over (\tilde{n}_0 + {Ns_2\over M+N})}
\ee
This is exactly the sum we found in (\ref{Ssum}), except with $M \leftrightarrow N$.  We thus know that to leading order it is given by
\be \label{eq:spans}
\tilde{S} \approx D_{s,s'}\pi \cot \Big{(}\pi {Ns_2 \over M+N}\Big{)} = -D_{s,s'}\pi\cot\Big{(}\pi{Ms_2\over M+N}\Big{)} = -S.
\ee

\subsection{The Full $\g^B$ for $s_2 \neq 0$}
Adding all the pieces together, we see that:
\be
\g^B_{ss'} = \g_{ss'}^{B(1)} + \g_{ss'}^{B(2)} = 2I
\ee
and while the near-pole contributions \eq{eq:sans} and \eq{eq:spans} do not individually vanish, they cancel.  The full answer is:
\be
\g^B_{ss'} = {1\over 2\pi(M+N)}{1\over s+s'}e^{-i(s+s')(\s_0-\t_0)}\mu_s\mu_{s'}
\ee
This agrees with the continuum limit form of $\g^B_{ss'}$ found in \cite{chmt}.

\subsection{Summary of results}
Here we summarize the results for the physical quantities $\g^B$ and $f^B$ in the continuum limit.  In both cases, the matrix is nonzero only when both indices are holomorphic or both indices are anti-holomorphic.  For $\g^B$, we find:
\be
\g^B_{ss'} = {1\over 2\pi(M+N)}{1\over s+s'}e^{-i(s+s')(\s_0-\t_0)}\mu_s\mu_{s'}
\ee
One can similarly check that the purely anti-holomorphic part is:
\be
\g^B_{\bar s \bar s'} = {1\over 2\pi(M+N)}{1\over \bar s + \bar s'}e^{i(\bar s + \bar s')(\s_0+\t_0)}\mu_{\bar s}\mu_{\bar s'} 
\ee
where we recall the notation
\be
\mu_s = 1 - e^{2\pi i M s} \,.
\label{mu2}
\ee
For $f^B$, we find:
\bea
f^{B(1)}_{qs} & = & (\a^{-1})^{(1)}_{sq} ~=~
\begin{cases}
\sqrt{M\over M+N}\d_{sq} & \qquad s_2 = 0 \\
-{1\over 2\pi i \sqrt{M(M+N)}}{1\over s-q}\m_{s}e^{-i(s-q)(\s_0-\t_0)} & \qquad s_2 \neq 0
\end{cases} \nn
f^{B(1)}_{\bar q \bar s} & = & (\a^{-1})^{(1)}_{\bar s \bar q} ~=~
\begin{cases}
\sqrt{M\over M+N}\d_{\bar s \bar q} & ~~~\qquad \bar s_2 = 0 \\
{1\over 2\pi i \sqrt{M(M+N)}}{1\over \bar s - \bar q}\mu_{-\bar s}e^{i(\bar s - \bar q)(\s_0+\t_0)} & ~~~\qquad \bar s_2 \neq 0
\end{cases} \nn
f^{B(2)}_{rs} & = & (\a^{-1})^{(2)}_{sr} ~=~
\begin{cases}
\sqrt{N\over M+N}\d_{sr} & ~~~\qquad s_2 = 0 \\
{1\over 2\pi i \sqrt{N(M+N)}}{1\over s-r}\m_{s}e^{-i(s-r)(\s_0-\t_0)} & ~~~\qquad s_2 \neq 0
\end{cases} \nn
f^{B(2)}_{\bar r \bar s} & = & (\a^{-1})^{(2)}_{\bar s \bar r} ~=~
\begin{cases}
\sqrt{N\over M+N}\d_{\bar s \bar r} & \qquad \bar s_2 = 0 \\
-{1\over 2\pi i \sqrt{N(M+N)}}{1\over \bar s - \bar r}\mu_{-\bar s}e^{i(\bar s - \bar r)(\s_0+\t_0)} & \qquad \bar s_2 \neq 0
\end{cases}
\eea
while all other components vanish.

\section{Discussion}\label{Discuss}

The D1D5 bound state is very interesting, because a complete understanding of  this system would capture the physics of near extremal black holes. But the D1D5 CFT is simple to study only at its free point - the `orbifold point'. To reach the coupling where black hole physics is captured, we must turn on the deformation operator with a finite strength. This deformation operator consists of a supercharge acting on a twist $\sigma_2$, so we must understand the dynamics of this twist. 

In earlier work it was recognized that the effect of the twist on the vacuum was to create a `squeezed state', described by coefficients $\gamma^B_{ss'}$. In the case of an initial excitation of a single mode, the state is also described by coefficients $f^B_{qs}$. We desire methods to compute these coefficients for the situation where we apply an arbitrary number of twist operators.  While the $\gamma^B_{ss'}$ and $f^B_{qs}$ were computed for the case of one twist insertion in \cite{acm2,acm3,chmt}, these computations are quite involved, and are not readily extendible to the situation with more twist insertions. Therefore in this paper we have developed a method of obtaining the $\gamma^B_{ss'}$ and $f^B_{qs}$ in an {\it approximation}, where the wavelengths of the excitations are short compared to the length of the `multiwound circle' on which the excitations live. Since the black hole states typically live on circles with very high winding \cite{maldasuss}, this approximation is expected to be very well satisfied for applications to black hole physics. We have termed this approximation the `continuum limit'. 

We recalled that the coefficients $\gamma^B_{ss'}$ are given by a relation of the schematic form   $\gamma^B\sim \alpha^{-1}\beta $, where $\alpha$ and $\beta$ are matrices of Bogoliubov coefficients created by the twist. We observed that the matrix $f^B$ is given by $f^B=(\a^{-1})^T$. We found and verified an expression for $\alpha^{-1}$ in the continuum limit, and we used this to calculate $\gamma^B$. 

In performing our computations, we found a 1-parameter family of $\{\alpha, \beta\}$ which lead to the {\it same } coefficients $\gamma^B$ and $f^B$. We have chosen to work with the $\alpha$ and $\beta$ which arise from a local definition of the twist as pictured in Fig.\;\ref{ftwo}. One may make a nonlocal change of variables, and go to a description where the $\alpha$ and $\beta$ are such that they connect left movers only to left movers and right movers only to right movers. Such $\alpha$ and $\beta$ for the case $M=N=1$  were used in \cite{ac}, and we have shown in Appendix \ref{NonUnique} that they lead to the same $\gamma^B$ and $f^B$ coefficients as the ones we compute with our choice of $\alpha$ and $\beta$. 

Let us comment on possible future applications of the methods developed in this paper. The coefficients $\alpha$ and $\beta$ are straightforward to compute for an arbitrary number of twists, without invoking any approximation. It is not clear how one would compute the inverse $\alpha^{-1}$ in general, since the matrix $\alpha$ is infinite dimensional; nor is it clear whether the matrix multiplication involved in $\alpha^{-1} \beta$ can be evaluated in general. However we have observed that for the case of one twist insertion, $\alpha^{-1}$ has a very simple form in the continuum limit. This leads us to be optimistic that one may be able to find $\alpha^{-1}$ in the continuum limit for multiple twist insertions. Further, for the case of one twist insertion we have seen that the matrix multiplication involved in $ \alpha^{-1}\beta$ becomes a simple contour integral; one may hope that a similar situation would hold for multiple twists. If this turns out to be the case, the techniques developed here may enable us to learn more about the nature of the thermalization process in the D1D5 CFT, and via holography, about black hole formation and evaporation.

\section*{Acknowledgements}

We thank Steve Avery, Borun Chowdhury, Amanda Peet and Ida Zadeh for discussions on various aspects of the twist interaction. The work is supported in part by DOE grant DE-FG02-91ER-40690.   

\begin{appendix}
\section{The Full $\a$, $\b$, and $\a^{-1}$ Matrices} \label{fullresults}
In this appendix we collect the full expressions for $\a$, $\b$, and $\a^{-1}$, first for $M=N=1$ and then for general $M$ and $N$.

\subsection{$M=N=1$}
When $M=N=1$, the full expressions for $\a$, $\b$, and $\a^{-1}$ are:
\bea
\alpha^{(1)}_{qs}&=&\begin{cases}
{1\over \sqrt{2}}\delta_{qs} & s \in \mathbb{Z} \\
{1 \over 2\pi i \sqrt{2}}{s+q \over \sqrt{sq}}{1\over s-q}e^{i(s-q)(\sigma_0-\tau_0)} & s \in \mathbb{Z}+\h
\end{cases}\nn
\a^{(1)}_{q\bar s}&=&\begin{cases}
0 & \bar s \in \mathbb{Z} \\
-{1 \over 2\pi i \sqrt{2}}{1 \over \sqrt{\bar s q}}e^{-i(\bar s + q)\s_0}e^{-i(\bar s - q)\t_0} & \bar s \in \mathbb{Z} + \h
\end{cases}\nn
\a^{(1)}_{\bar q s}&=&\begin{cases}
0 & s \in \mathbb{Z} \\
{1 \over 2\pi i \sqrt{2}}{1 \over \sqrt{s \bar q}}e^{i(s + \bar q)\s_0}e^{-i(s - \bar q)\t_0} & s \in \mathbb{Z} + \h
\end{cases}\nn
\alpha^{(1)}_{\bar q \bar s}&=&\begin{cases}
{1\over \sqrt{2}}\delta_{\bar q \bar s} & \bar s \in \mathbb{Z} \\
-{1 \over 2\pi i \sqrt{2}}{\bar s + \bar q \over \sqrt{\bar s \bar q}}{1\over \bar s - \bar q}e^{-i(\bar s - \bar q )(\sigma_0+\tau_0)} & \bar s \in \mathbb{Z}+\h
\end{cases}
\eea

\bea
\beta^{(1)}_{qs}&=&\begin{cases}
0 & s \in \mathbb{Z} \\
{1 \over 2\pi i \sqrt{2}}{s-q \over \sqrt{sq}}{1\over s+q}e^{-i(s+q)(\sigma_0-\tau_0)} & s \in \mathbb{Z} + \h
\end{cases}\nn
\beta^{(1)}_{q\bar s}&=&\begin{cases}
0 & \bar s \in \mathbb{Z} \\
-{1 \over 2\pi i \sqrt{2}}{1 \over \sqrt{\bar s q}}e^{i(\bar s - q)\s_0}e^{i(\bar s + q)\t_0} & \bar s \in \mathbb{Z} + \h
\end{cases}\nn
\b^{(1)}_{\bar q s}&=&\begin{cases}
0 & s \in \mathbb{Z} \\
{1 \over 2\pi i \sqrt{2}}{1 \over \sqrt{s \bar q}}e^{-i(s - \bar q)\s_0}e^{i(s + \bar q)\t_0} & s \in \mathbb{Z} + \h
\end{cases}\nn
\b^{(1)}_{\bar q \bar s}&=&\begin{cases}
0 & \bar s \in \mathbb{Z} \\
-{1 \over 2\pi i \sqrt{2}}{\bar s - \bar q \over \sqrt{\bar s \bar q}}{1\over \bar s + \bar q}e^{i(\bar s + \bar q )(\sigma_0+\tau_0)} & \bar s \in \mathbb{Z}+\h
\end{cases}
\eea

\bea
\left (\alpha^{-1}\right)^{(1)} _{sq}&=& \begin{cases}
{1\over \sqrt{2}}\delta_{sq} & s \in \mathbb{Z} \\
-{1\over \pi i\sqrt{2}}{1\over s-q} \, e^{-i(s-q)(\sigma_0-\tau_0)} & s \in \mathbb{Z}+\h
\end{cases}\nn
\left (\alpha^{-1}\right)^{(1)} _{\bar s \bar q}&=& \begin{cases}
{1\over \sqrt{2}}\delta_{\bar s \bar q} & \bar s \in \mathbb{Z} \\
{1\over \pi i\sqrt{2}}{1\over \bar s - \bar q} \, e^{i(\bar s - \bar q)(\sigma_0+\tau_0)} & \bar s \in \mathbb{Z}+\h
\end{cases}\nn
\left (\alpha^{-1}\right)^{(1)} _{\bar s q}&=&\left (\alpha^{-1}\right)^{(1)} _{s \bar q}~=~0
\eea

\bea
\alpha^{(2)}_{rs}&=& \begin{cases}
{1\over \sqrt{2}}\delta_{rs} & s\in \mathbb{Z} \\
-{1 \over 2\pi i \sqrt{2}}{s+r \over \sqrt{sr}}{1\over s-r}e^{i(s-r)(\sigma_0-\tau_0)} & s\in \mathbb{Z}+\h
\end{cases}\nn
\a^{(2)}_{r\bar s}&=&\begin{cases}
0 & \bar s \in \mathbb{Z} \\
{1 \over 2\pi i \sqrt{2}}{1 \over \sqrt{\bar s r}}e^{-i(\bar s + r)\s_0}e^{-i(\bar s - r)\t_0} & \bar s \in \mathbb{Z} + \h
\end{cases}\nn
\a^{(2)}_{\bar r s}&=&\begin{cases}
0 & s \in \mathbb{Z} \\
-{1 \over 2\pi i \sqrt{2}}{1 \over \sqrt{s \bar r}}e^{i(s + \bar r)\s_0}e^{-i(s - \bar r)\t_0} & s \in \mathbb{Z} + \h
\end{cases}\nn
\alpha^{(2)}_{\bar r \bar s}&=&\begin{cases}
{1\over \sqrt{2}}\delta_{\bar r \bar s} & \bar s \in \mathbb{Z} \\
{1 \over 2\pi i \sqrt{2}}{\bar s + \bar r \over \sqrt{\bar s \bar r}}{1\over \bar s - \bar r}e^{-i(\bar s - \bar r )(\sigma_0+\tau_0)} & \bar s \in \mathbb{Z}+\h
\end{cases}
\eea

\bea
\beta^{(2)}_{rs}&=&\begin{cases}
0 & s \in \mathbb{Z} \\
-{1 \over 2\pi i \sqrt{2}}{s-r \over \sqrt{sr}}{1\over s+r}e^{-i(s+r)(\sigma_0-\tau_0)}  & s \in \mathbb{Z} + \h
\end{cases}\nn
\b^{(2)}_{r\bar s}&=&\begin{cases}
0 & \bar s \in \mathbb{Z} \\
{1 \over 2\pi i \sqrt{2}}{1 \over \sqrt{\bar s r}}e^{i(\bar s - r)\s_0}e^{i(\bar s + r)\t_0} & \bar s \in \mathbb{Z} + \h
\end{cases}\nn
\b^{(2)}_{\bar r s}&=&\begin{cases}
0 & s \in \mathbb{Z} \\
-{1 \over 2\pi i \sqrt{2}}{1 \over \sqrt{s \bar r}}e^{-i(s - \bar r)\s_0}e^{i(s + \bar r)\t_0} & s \in \mathbb{Z} + \h
\end{cases}\nn
\b^{(2)}_{\bar r \bar s}&=&\begin{cases}
0 & \bar s \in \mathbb{Z} \\
{1 \over 2\pi i \sqrt{2}}{\bar s - \bar r \over \sqrt{\bar s \bar r}}{1\over \bar s + \bar r}e^{i(\bar s + \bar r )(\sigma_0+\tau_0)} & \bar s \in \mathbb{Z}+\h
\end{cases}
\eea

\bea
\left (\alpha^{-1}\right)^{(2)} _{sr}&=& \begin{cases}
{1\over \sqrt{2}}\delta_{sr} & s \in \mathbb{Z} \\
{1\over \pi i \sqrt{2}}{1\over s-r} \, e^{-i(s-r)(\sigma_0-\tau_0)} & s \in \mathbb{Z}+\h
\end{cases}\nn
\left (\alpha^{-1}\right)^{(2)} _{\bar s \bar r}&=& \begin{cases}
{1\over \sqrt{2}}\delta_{\bar s \bar r} & \bar s \in \mathbb{Z} \\
-{1\over \pi i \sqrt{2}}{1\over \bar s - \bar r} \, e^{i(\bar s - \bar r)(\sigma_0+\tau_0)} & \bar s \in \mathbb{Z}+\h
\end{cases}\nn
\left (\alpha^{-1}\right)^{(2)} _{\bar s r}&=&\left (\alpha^{-1}\right)^{(2)} _{s \bar r}~=~0
\eea

\subsection{General $M$ and $N$}
For general $M$ and $N$, the full expressions for $\a$, $\b$, and $\a^{-1}$ are:
\bea
\a^{(1)}_{qs} & = &
\begin{cases}
\sqrt{M\over M+N}\d_{qs} & s_2 = 0 \\
{1 \over 4\pi i \sqrt{M(M+N)}}{s+q \over \sqrt{sq}}{1\over s-q}\mu_{-s}e^{i(s-q)(\s_0-\t_0)} & s_2 \neq 0
\end{cases} \nn
\a^{(1)}_{q \bar s} & = &-{1 \over 4\pi i \sqrt{M(M+N)}}{1\over\sqrt{\bar s q}}\mu_{\bar s}e^{-i(\bar s+ q)\s_0} e^{-i(\bar s - q) \t_0} \nn
\a^{(1)}_{\bar q s} & = &{1 \over 4\pi i \sqrt{M(M+N)}}{1\over\sqrt{s \bar q}}\mu_{-s}e^{i(s+ \bar q)\s_0} e^{-i(s - \bar q)\t_0} \nn
\a^{(1)}_{\bar q \bar s} & = &
\begin{cases}
\sqrt{M\over M+N}\d_{\bar q \bar s} & \bar s_2 = 0 \\
-{1 \over 4\pi i \sqrt{M(M+N)}}{\bar s + \bar q \over \sqrt{\bar s \bar q}}{1\over \bar s - \bar q}\mu_{\bar s}e^{-i(\bar s - \bar q)(\s_0+\t_0)} & \bar s_2 \neq 0
\end{cases}
\eea

\bea
\b^{(1)}_{qs} &=& {1 \over 4\pi i \sqrt{M(M+N)}}{s-q \over \sqrt{sq}}{1\over s+q}\mu_se^{-i(s+q)(\s_0-\t_0)} \nn
\b^{(1)}_{q \bar s} & = &-{1 \over 4\pi i \sqrt{M(M+N)}}{1\over\sqrt{\bar s q}}\mu_{-\bar s}e^{i(\bar s- q)\s_0}e^{i(\bar s +  q)\t_0} \nn
\b^{(1)}_{\bar q s} & = &{1 \over 4\pi i \sqrt{M(M+N)}}{1\over\sqrt{s \bar q}}\mu_{ s}e^{-i(s- \bar q)\s_0}e^{i(s + \bar q)\t_0} \nn
\b^{(1)}_{\bar q \bar s} & = &-{1 \over 4\pi i \sqrt{M(M+N)}}{\bar s - \bar q \over \sqrt{\bar s \bar q}}{1\over \bar s + \bar q}\mu_{- \bar s}e^{i(\bar s + \bar q)(\s_0+\t_0)}
\eea

\bea
(\a^{-1})^{(1)}_{sq} & = &
\begin{cases}
\sqrt{M\over M+N}\d_{sq} & s_2 = 0 \\
-{1\over 2\pi i \sqrt{M(M+N)}}{1\over s-q}\m_{s}e^{-i(s-q)(\s_0-\t_0)} & s_2 \neq 0
\end{cases} \nn
(\a^{-1})^{(1)}_{\bar s \bar q} & = &
\begin{cases}
\sqrt{M\over M+N}\d_{\bar s \bar q} & \bar s_2 = 0 \\
{1\over 2\pi i \sqrt{M(M+N)}}{1\over \bar s - \bar q}\mu_{-\bar s}e^{i(\bar s - \bar q)(\s_0+\t_0)} & \bar s_2 \neq 0
\end{cases} \nn
(\a^{-1})^{(1)}_{\bar s q} & = & (\a^{-1})^{(1)}_{s \bar q} ~=~ 0
\eea

\bea
\a^{(2)}_{rs} & = &
\begin{cases}
\sqrt{N\over M+N}\d_{rs} & s_2 = 0 \\
-{1 \over 4\pi i \sqrt{N(M+N)}}{s+r \over \sqrt{sr}}{1\over s-r}\mu_{-s}e^{i(s-r)(\s_0-\t_0)} & s_2 \neq 0
\end{cases} \nn
\a^{(2)}_{r \bar s} & = &{1 \over 4\pi i \sqrt{N(M+N)}}{1\over\sqrt{\bar s r}}\mu_{\bar s}e^{-i(\bar s+ r)\s_0} e^{-i(\bar s - r) \t_0} \nn
\a^{(2)}_{\bar r s} & = &-{1 \over 4\pi i \sqrt{N(M+N)}}{1\over\sqrt{s \bar r}}\mu_{-s}e^{i(s+ \bar r)\s_0} e^{-i(s - \bar r)\t_0} \nn
\a^{(2)}_{\bar r \bar s} & = &
\begin{cases}
\sqrt{N\over M+N}\d_{\bar r \bar s} & \bar s_2 = 0 \\
{1 \over 4\pi i \sqrt{N(M+N)}}{\bar s + \bar r \over \sqrt{\bar s \bar r}}{1\over \bar s - \bar r}\mu_{\bar s}e^{-i(\bar s - \bar r) (\s_0+\t_0)} & \bar s_2 \neq 0
\end{cases}
\eea

\bea
\b^{(2)}_{rs} &=& -{1 \over 4\pi i \sqrt{N(M+N)}}{s-r \over \sqrt{sr}}{1\over s+r}\mu_se^{-i(s+r)(\s_0-\t_0)} \nn
\b^{(2)}_{r \bar s} & = &{1 \over 4\pi i \sqrt{N(M+N)}}{1\over\sqrt{\bar s r}}\mu_{-\bar s}e^{i(\bar s- r)\s_0}e^{i(\bar s +  r)\t_0} \nn
\b^{(2)}_{\bar r s} & = &-{1 \over 4\pi i \sqrt{N(M+N)}}{1\over\sqrt{s \bar r}}\mu_{ s}e^{-i(s- \bar r)\s_0}e^{i(s + \bar q)\t_0} \nn
\b^{(2)}_{\bar r \bar s} & = &{1 \over 4\pi i \sqrt{N(M+N)}}{\bar s - \bar r \over \sqrt{\bar s \bar r}}{1\over \bar s + \bar r}\mu_{- \bar s}e^{i(\bar s + \bar r)(\s_0+\t_0)}
\eea

\bea
(\a^{-1})^{(2)}_{sr} & = &
\begin{cases}
\sqrt{N\over M+N}\d_{sr} & s_2 = 0 \\
{1\over 2\pi i \sqrt{N(M+N)}}{1\over s-r}\m_{s}e^{-i(s-r)(\s_0-\t_0)} & s_2 \neq 0
\end{cases} \nn
(\a^{-1})^{(2)}_{\bar s \bar r} & = &
\begin{cases}
\sqrt{N\over M+N}\d_{\bar s \bar r} & \bar s_2 = 0 \\
-{1\over 2\pi i \sqrt{N(M+N)}}{1\over \bar s - \bar r}\mu_{-\bar s}e^{i(\bar s - \bar r)(\s_0+\t_0)} & \bar s_2 \neq 0
\end{cases} \nn
(\a^{-1})^{(2)}_{\bar s r} & = & (\a^{-1})^{(2)}_{r \bar q} ~=~ 0
\eea

\newpage

\section{Ambiguity in the choice of $\alpha, \beta$} \label{NonUnique}

In this appendix we comment on the relation of our work to that of \cite{ac}, which deals with $M=N=1$.  The $\alpha$ matrices we have computed for $M=N=1$ are given in Appendix \ref{fullresults}. In \cite{ac}, however, it was argued that the following could be used instead:
\bea
\tilde \alpha^{(1)}_{qs}&=&\begin{cases}
{1\over \sqrt{2}}\delta_{qs} & s \in \mathbb{Z} \\
{1 \over \pi i \sqrt{2}}{\sqrt{s\over q}}{1\over s-q}e^{i(s-q)(\sigma_0-\tau_0)} & s \in \mathbb{Z}+\h
\end{cases}\nn
\tilde \beta^{(1)}_{qs}&=&\begin{cases}
0 & s \in \mathbb{Z} \\
{1 \over \pi i \sqrt{2}}{\sqrt{s\over q}}{1\over s+q}e^{-i(s+q)(\sigma_0-\tau_0)} & s \in \mathbb{Z} + \h
\end{cases}
\eea
\bea
\tilde \alpha^{(2)}_{rs}&=&\begin{cases}
{1\over \sqrt{2}}\delta_{rs} & s \in \mathbb{Z} \\
-{1 \over \pi i \sqrt{2}}{\sqrt{s\over r}}{1\over s-r}e^{i(s-r)(\sigma_0-\tau_0)} & s \in \mathbb{Z}+\h
\end{cases}\nn
\tilde \beta^{(2)}_{rs}&=&\begin{cases}
0 & s \in \mathbb{Z} \\
-{1 \over \pi i \sqrt{2}}{\sqrt{s\over r}}{1\over s+r}e^{-i(s+r)(\sigma_0-\tau_0)} & s \in \mathbb{Z} + \h
\end{cases}
\eea
and there are no terms which involve both left and right movers: for example $\tilde \alpha^{(1)}_{q\bar s}=0$. The corresponding anti-holomorphic expressions are:
\bea
\tilde \alpha^{(1)}_{\bar q \bar s}&=&\begin{cases}
{1\over \sqrt{2}}\delta_{\bar q \bar s} & \bar s \in \mathbb{Z} \\
-{1 \over \pi i \sqrt{2}}{\sqrt{\bar s\over \bar q}}{1\over \bar s - \bar q}e^{-i(\bar s - \bar q)(\sigma_0+\tau_0)} & \bar s \in \mathbb{Z}+\h
\end{cases}\nn
\tilde \beta^{(1)}_{\bar q \bar s}&=&\begin{cases}
0 & \bar s \in \mathbb{Z} \\
-{1 \over \pi i \sqrt{2}}{\sqrt{\bar s\over \bar q}}{1\over \bar s + \bar q}e^{i(s+q)(\sigma_0+\tau_0)} & \bar s \in \mathbb{Z} + \h
\end{cases}
\eea
\bea
\tilde \alpha^{(2)}_{\bar r \bar s}&=&\begin{cases}
{1\over \sqrt{2}}\delta_{\bar r \bar s} & \bar s \in \mathbb{Z} \\
{1 \over \pi i \sqrt{2}}{\sqrt{\bar s\over \bar r}}{1\over \bar s-\bar r}e^{-i(\bar s-\bar r)(\sigma_0+\tau_0)} & \bar s \in \mathbb{Z}+\h
\end{cases}\nn
\tilde \beta^{(2)}_{\bar r \bar s}&=&\begin{cases}
0 & \bar s \in \mathbb{Z} \\
{1 \over \pi i \sqrt{2}}{\sqrt{\bar s\over \bar r}}{1\over \bar s+\bar r}e^{i(\bar s+\bar r)(\sigma_0+\tau_0)} & s \in \mathbb{Z} + \h
\end{cases}
\eea
Let us write
\be
\Delta \alpha=\alpha-\tilde\alpha, ~~~~\Delta \beta=\beta-\tilde\beta
\ee
Then we find that $\D \a, \D \beta$ vanish when $s$ (or $\bar s$) is an integer. When $s$ (or $\bar s$) is half-integer, we have:
\bea
\D\alpha^{(1)}_{qs}&=&-{1 \over 2\pi i \sqrt{2}}{1 \over \sqrt{sq}}e^{i(s-q)(\sigma_0-\tau_0)}\nn
\D\a^{(1)}_{q\bar s}&=&-{1 \over 2\pi i \sqrt{2}}{1 \over \sqrt{\bar s q}}e^{-i(\bar s + q)\s_0}e^{-i(\bar s - q)\t_0} \nn
\D\a^{(1)}_{\bar q s}&=&{1 \over 2\pi i \sqrt{2}}{1 \over \sqrt{s \bar q}}e^{i(s + \bar q)\s_0}e^{-i(s - \bar q)\t_0} \nn
\D\alpha^{(1)}_{\bar q \bar s}&=&{1 \over 2\pi i \sqrt{2}}{1 \over \sqrt{\bar s \bar q}}e^{-i(\bar s - \bar q )(\sigma_0+\tau_0)}
\eea
\bea
\D\beta^{(1)}_{qs}&=&-{1 \over 2\pi i \sqrt{2}}{1 \over \sqrt{sq}}e^{-i(s+q)(\sigma_0-\tau_0)}\nn
\D\beta^{(1)}_{q\bar s}&=&-{1 \over 2\pi i \sqrt{2}}{1 \over \sqrt{\bar s q}}e^{i(\bar s - q)\s_0}e^{i(\bar s + q)\t_0}\nn
\D\b^{(1)}_{\bar q s}&=&{1 \over 2\pi i \sqrt{2}}{1 \over \sqrt{s \bar q}}e^{-i(s - \bar q)\s_0}e^{i(s + \bar q)\t_0}\nn
\D\b^{(1)}_{\bar q \bar s}&=&{1 \over 2\pi i \sqrt{2}}{1 \over \sqrt{\bar s \bar q}}e^{i(\bar s + \bar q )(\sigma_0+\tau_0)}
\eea
and
\bea
\D\alpha^{(2)}_{rs}&=&{1 \over 2\pi i \sqrt{2}}{1 \over \sqrt{sr}}e^{i(s-r)(\sigma_0-\tau_0)}\nn
\D\a^{(2)}_{r\bar s}&=&{1 \over 2\pi i \sqrt{2}}{1 \over \sqrt{\bar s r}}e^{-i(\bar s + r)\s_0}e^{-i(\bar s - r)\t_0}\nn
\D\a^{(2)}_{\bar r s}&=&-{1 \over 2\pi i \sqrt{2}}{1 \over \sqrt{s \bar r}}e^{i(s + \bar r)\s_0}e^{-i(s - \bar r)\t_0}\nn
\D\alpha^{(2)}_{\bar r \bar s}&=&-{1 \over 2\pi i \sqrt{2}}{1 \over \sqrt{\bar s \bar r}}e^{-i(\bar s - \bar r )(\sigma_0+\tau_0)}
\eea
\bea
\D\beta^{(2)}_{rs}&=&{1 \over 2\pi i \sqrt{2}}{1 \over \sqrt{sr}}e^{-i(s+r)(\sigma_0-\tau_0)}\nn
\D\b^{(2)}_{r\bar s}&=&{1 \over 2\pi i \sqrt{2}}{1 \over \sqrt{\bar s r}}e^{i(\bar s - r)\s_0}e^{i(\bar s + r)\t_0}\nn
\D\b^{(2)}_{\bar r s}&=&-{1 \over 2\pi i \sqrt{2}}{1 \over \sqrt{s \bar r}}e^{-i(s - \bar r)\s_0}e^{i(s + \bar r)\t_0}\nn
\D\b^{(1)}_{\bar r \bar s}&=&-{1 \over 2\pi i \sqrt{2}}{1 \over \sqrt{\bar s \bar r}}e^{i(\bar s + \bar r )(\sigma_0+\tau_0)}\,.
\eea

Using methods similar to those we used in Section \ref{1,1twist}, one can show that at order $k^0$:
\bea
\alpha^{-1} \Delta\alpha&=&0 \label{jfivep}\\
(\Delta\alpha) \alpha^{-1}&=&0
\label{jfive}
\eea
We now see that $\alpha$ and $\tilde\alpha$ have the {\it same} inverse $\alpha^{-1}$.  In fact, we can make an even stronger statement.  If we take the exact form of $\a^{-1}$ computed in \cite{acm2,chmt}, the relations (\ref{jfivep}) and (\ref{jfive}) hold exactly, after including also the contribution from zero modes\footnote{In addition, we have verified numerically in a number of examples that the exact $f^B$ for general $M$, $N$ computed in \cite{chmt} exactly satisfies $(f^B)^T\alpha = \alpha(f^B)^T= \one$, after including zero modes.}.  Thus, the non-uniqueness of $\a$ and $\b$ is not an artifact of the continuum limit.  It is a result of the fact that we are dealing with infinite matrix multiplication.

One can further show that:
\bea
\alpha^{-1} \Delta\beta&=&0\nn
(\Delta\beta) \alpha^{-1}&=&0
\eea
Thus we find that
\be
\alpha^{-1}\tilde \beta = \alpha^{-1}\beta =\gamma^B
\ee
leaving $\g^B$ unique, as expected.  Thus the physically relevant quantities $\gamma^B$ and $f=\left (\alpha^{-1}\right )^T$ are the same for the Bogoliubov coefficients $\alpha,\beta $ and the coefficients $\tilde\alpha, \tilde\beta$. In fact we have a 1-parameter family of Bogoliubov coefficients
\be
\alpha_\nu=\alpha+\nu \Delta\alpha, ~~~~\beta_\nu=\beta+\nu \Delta \beta
\ee
which all give the same values of the physically relevant  quantities $\gamma^B$ and $f=\left (\alpha^{-1}\right )^T$. 

Of course, in mathematics it has long been known that infinite matrix multiplication is not in general associative, and that left and right inverses do not always exist and are not always unique (see e.g.~\cite{infinitematrices}).  However, such subtleties are not usually encountered in field theory.  To illustrate how non-associativity arises in the present case, consider the triple  product
\be
\sum_{s,q'}\a^{(1)}_{qs}\left (\a^{-1}\right )^{(1)}_{sq'}\D\a^{(1)}_{q's'} \,.
\ee

There are two conflicting relations one can use here.  The first is (\ref{jfivep}), which would imply that the result is 0.  On the other hand, we have:
\be\label{identity}
\a\a^{-1} = \one
\ee
which would imply that the result is $\D\a^{(1)}_{qs'}$.  This discrepancy arises from the fact that (\ref{jfivep}) only holds if the index $q'$ is allowed to span a region that includes $q' \gg s$, while (\ref{identity}) only holds if the index $s$ is allowed to span a region that includes $s \gg q'$.  Thus the order in which we perform the $s$ and $q'$ sums is important, which is to say that the matrix multiplication is not associative.  If the $q'$ sum is performed first, (\ref{jfivep}) applies but (\ref{identity}) does not.  The reverse holds if the $s$ sum is performed first.

Finally, we note that ordering ambiguities involving infinite sums were also encountered in \cite{ac}.  It is possible that there is some relation between those ordering ambiguities and the non-associativity observed here.

\section{Verifying $\a^{-1}$ for General $M$ and $N$}\label{InverseCheck}

In this appendix we verify that our expression for $\a^{-1}$ for general $M$ and $N$ satisfies, at leading order in $Mk$ $(\sim Nk)$:
\be
\a\a^{-1} ~=~ \a^{-1}\a ~=~ \one \,.
\ee
We will continue to mostly leave the higher order corrections implicit, and write only the order $(Mk)^0$ terms.

\subsection{Diagonal Blocks}
We first show that:
\be
\sum_s \a^{(1)}_{qs}\left ( \a^{-1}\right )^{(1)}_{sq'} = \d_{qq'}
\ee
The $s_2 = 0$ terms for $\a^{(1)}_{qs}$ and $\left ( \a^{-1}\right )^{(1)}_{sq'}$ can only occur when $q_2$ and $q'_2$ are also zero.  We thus have:
\bea
\sum\limits_s \a^{(1)}_{qs}(\a^{-1})^{(1)}_{sq'} & = & {M \over M+N}\d_{qq'}\d_{q_2,0}\\
&&{}+ {e^{-i(q-q')(\s_0-\t_0)}\over 8\pi^2 M(M+N)}\sum_{s>0,\,s_2 \neq 0}|1-e^{2\pi i Ms}|^2{s+q \over \sqrt{sq}}{1\over s-q}{1\over s-q'} \nonumber
\eea
When $q = q'$, the sum simplifies.  However, when $q \neq q'$, the sum is asymmetric around the poles, so we will have to treat those regions carefully.  We consider $q=q'$ first.

\subsubsection{$q = q'$}\label{diagonalblocksnonzero}
When $q=q'$, we have:
\bea \label{sumone}
\sum\limits_s \a^{(1)}_{qs}(\a^{-1})^{(1)}_{sq} & = & {M \over M+N}\d_{q_2,0}\\
&&{}+ {1\over 8\pi^2 M(M+N)}\sum_{s>0,\,s_2 \neq 0}|1-e^{2\pi i Ms}|^2{s+q \over \sqrt{sq}}{1\over (s-q)^2} \nonumber
\eea
The $(s-q)^2$ term in the denominator gives a sharp peak, and since each term in the sum is non-negative, there is no cancellation in the peak region.  If the sum extends well beyond this peak, we can replace the lower bound of the sum by $-\infty$ with only subleading corrections.  Let us check that this is the case.

First, recall that the spacing of elements in the sum is for the most part $\D s = \tfrac{1}{M+N}$ (though occasionally it is twice this as we skip over an element with $s_2 = 0$).  This means that the largest terms in the sum have $\tfrac{1}{s-q} \sim M+N$ or larger.  Such terms will thus be at least of order $\tfrac{M+N}{M}$.  Let us compare this to the size of the terms at the lower range of summation.  The lowest value of $s$ is $\tfrac{1}{M+N} \ll q$, and so such terms have magnitude of order $(M^2(M+N)q^3)^{-1/2}$.  Since $Mq \gg 1$, these terms are strongly suppressed.  In other words, our sum does extend well beyond the peak, and we can safely extend its range to $-\infty$ with only subleading corrections\footnote{Quantitatively, the corrections are of order $(M(M+N)Mq)^{-1/2} \ll \tfrac{M+N}{M}$.}.  Furthermore, we can use the peaked nature of the sum to approximate $s \approx q$:
\be
{s+q \over \sqrt{sq}} \approx 2
\ee
We then have:
\bea
\sum_s \a^{(1)}_{qs}(\a^{-1})^{(1)}_{sq} & \approx & {M \over M+N}\d_{q_2,0}+ {1\over 4\pi^2 M(M+N)}\sum_{s>0,\,s_2 \neq 0}|1-e^{2\pi i Ms}|^2{1\over (s-q)^2} \nn
&\approx& {M \over M+N}\d_{q_2,0} + \sum_{s_2 = 1}^{m+n-1}{|1-e^{2\pi i {M\over M+N}s_2}|^2 \over 4\pi^2 M(M+N)}\nn
&&\times  \sum_{s_1 = 0}^{Y-1}\sum_{n_0=-\infty}^{\infty}{1\over (n_0 + {s_1 - q_1 \over Y} + {s_2 \over M+N} - {q_2 \over M})^2}
\eea
Where $n_0 = s_0 - q_0$, and we used (\ref{mu}).  The sum over $n_0$ can be evaluated analytically:
\be
\sum_{n_0 = -\infty}^{\infty} {1\over (n_0 + a)^2} = {\pi^2 \over \sin^2(\pi a)}
\ee
We can also write the mod squared term in terms of a cosine:
\be
\left | 1-e^{2\pi i {M\over M+N}s_2}\right |^2 = 2\left ( 1 - \cos \left ( 2\pi{M\over M+N}s_2\right ) \right )
\ee
We are then left with:
\bea
\sum_s \a^{(1)}_{qs}(\a^{-1})^{(1)}_{sq} & \approx & {M \over M+N}\d_{q_2,0} \\
&&{}+ {1\over 2M(M+N)}\sum_{s_2 = 1}^{m+n-1}\sum_{s_1 = 0}^{Y-1}{1-\cos(2\pi {M\over M+N}s_2) \over \sin^2\big{(} \pi ({s_1 - q_1 \over Y}+ {s_2 \over M+N} - {q_2 \over M})\big{)}}\nonumber
\eea
The sum over $s_1$ and the periodicity of the sine squared function means that the result is clearly independent of the value $q_1$.  This double sum could not be handled analytically.  However, for a wide range of chosen values for $M$ and $N$, we find:
\be
\sum_{s_2 = 1}^{m+n-1}\sum_{s_1 = 0}^{Y-1}{1-\cos(2\pi {M\over M+N}s_2) \over \sin^2\big{(} \pi ({s_1 - q_1 \over Y}+ {s_2 \over M+N} - {q_2 \over M})\big{)}} = \begin{cases}
2MN & q_2 = 0 \\
2M(M+N) & q_2 \in \{1,2,\ldots m-1\}
\end{cases}
\ee
Then for $q_2 = 0$, we have:
\be
\sum_s \a^{(1)}_{qs}(\a^{-1})^{(1)}_{sq} \approx {M \over M+N} + {2MN \over 2M(M+N)} = 1
\ee
And for $q_2 \neq 0$:
\be
\sum_s \a^{(1)}_{qs}(\a^{-1})^{(1)}_{sq} \approx {2M(M+N) \over 2M(M+N)} = 1.
\ee

\subsubsection{$q \neq q'$} \label{diagonalblockoffdiagonalelements}
When $q \neq q'$, the $s_2 = 0$ term automatically vanishes.  We then have:
\be\label{qneqq'sum}
\sum_s \a^{(1)}_{qs}(\a^{-1})^{(1)}_{sq'} = e^{-i(q-q')(\s_0-\t_0)}\sum_{s>0,\,s_2 \neq 0}{|1-e^{2\pi i Ms}|^2\over 8\pi^2 M(M+N)}{s+q \over \sqrt{sq}}{1\over s-q}{1\over s-q'}
\ee
Far from $q$ and $q'$, we can treat the sum as a principal value integral.  But we cannot do this near $q$ or $q'$, so we cut out boxes around these poles of size $L = \e k$ as done in (\ref{range}).  We make the boxes symmetric around the values of $s$ which minimize the distances $|s-q|$ and $|s-q'|$.  In these regions, the momenta $q$, $q'$, and $s$ will all be of the same order of magnitude $k$, so to leading order we have:
\be
{s+q \over \sqrt{sq}} \approx 2 \label{approx}
\ee

Let us split the poles apart into partial fractions:
\be
{1 \over (s-q)(s-q')}={1 \over (q-q')}\left ({1\over (s-q)}-{1\over (s-q')} \right )\label{partialfractions}
\ee
In order for the continuum limit to remain a good approximation, we must be able to treat the modes between $q$ and $q'$ as a continuum.  For this, we need the spacing between $q$ and $q'$ to be much larger than the spacing between each mode, i.e. we require
\be
q-q' \gg {1\over M}\,.
\ee
Let us define $\tilde s$ to be the value of $s$ which minimizes $|s-q|$, and $\tilde s'$ to be the value of $s$ which minimizes $|s-q'|$. 
We then apply (\ref{partialfractions}) and (\ref{approx}) to (\ref{qneqq'sum}), obtaining:
\be
\sum_s \a^{(1)}_{qs}(\a^{-1})^{(1)}_{sq'} = e^{-i(q-q')(\s_0-\t_0)}\left ( Q + Q' + \sum_{s_2=1}^{m+n-1}|1-e^{2\pi i Ms}|^2\sum_{s_1=0}^{Y-1}I \right ) \nonumber
\ee
where
\bea \label{sumpieces}
Q&=& {1\over 4\pi^2 M(M+N)}{1\over (q-q')}\sum_{s = \tilde s -L}^{\tilde s +L}{|1-e^{2\pi i Ms}|^2\over (s-q)} \nn
Q'&=& -{1\over 4\pi^2 M(M+N)}{1\over (q-q')}\sum_{s=\tilde s'-L}^{\tilde s'+L}{|1-e^{2\pi i Ms}|^2\over (s-q')} \nn
I & = & {1\over 8\pi^2 M(M+N)}{\cal P}\ints_0^{\infty}{s+q \over \sqrt{sq}}{1\over s-q}{1\over s-q'}\diff s_0 \,.
\eea

Let us tackle the integral first.  For each fixed $s_1$ and $s_2$, we have $\diff s_0 = \diff s$.  We perform the transformation $\tilde{s}^2 = s$, for which $\diff s = 2\tilde{s} \diff \tilde{s}$, and use the evenness of the integrand to extend the integral along the entire real axis.  Closing in the upper half plane we have:
\bea
I & = & {\cal P}\ints_{-\infty}^{\infty}{\tilde{s}^2+q \over \sqrt{q}}{1\over \tilde{s}^2-q}{1\over \tilde{s}^2-q'}\diff \tilde{s} \nn
&=& {2\pi i \over 2} \Big{(} {2q \over -2q(q-q')} + {2q \over 2q(q-q')} + {q'+q \over -2\sqrt{qq'}(q'-q)} +  {q'+q \over 2\sqrt{qq'}(q'-q)} \Big{)} \nn
&=& 0 \,.
\eea

Now let us turn to $Q$:
\be \label{eq:qsum}
Q = {1\over 4\pi^2 M(M+N)}{1\over q-q'}\sum_{s = \tilde s - L}^{\tilde s + L}{|1-e^{2\pi i Ms}|^2\over s-q}
\ee
When $q_2 = 0$, the summand is antisymmetric around $q$ while the bounds are symmetric, so we have $Q=0$.  But when $q_2 \neq 0$, the summand is not perfectly antisymmetric, and so a priori the cancellation need not be exact.  Let us therefore check that $Q$ is indeed subleading relative to the $q=q'$ entries of $\a\a^{-1}$.

In the expression (\ref{eq:qsum}), it would be convenient if we could extend the bounds of the sum to $\pm \infty$.  To see that we can indeed do this without changing the leading order behavior, let us examine the terms at the edge of our box, with $s = \tilde s \pm L$.  Defining $\d = \tilde s - q$, these two terms together give a contribution:
\be
Q_{\pm L} = {1\over 2\pi^2 M(M+N)}{1\over q-q'}\left ({1-\cos\left ( 2\pi M (\d+L)\right )\over \d + L} + {1-\cos\left ( 2\pi M (\d-L)\right )\over \d - L}\right )
\ee
Note that $\d \leq \D s \ll L$, so the denominators of these two terms are approximately equal in magnitude.  We thus write:
\bea\label{edgecontribution}
Q_{\pm L} &\approx& {1\over 2\pi^2 M(M+N)}{1\over q-q'}{\cos\left ( 2\pi M (\d-L)\right )-\cos\left ( 2\pi M (\d+L)\right )\over L}\nn
&=&{1\over \pi^2 M(M+N)}{1\over q-q'}\sin(2\pi M \d){\sin(2\pi M L)\over L}\nn
&=&{1\over \pi^2 M}{1\over q-q'}\sin(2\pi M \d){\sin(2\pi {MK\over M+N})\over K}
\eea
where $K = \tfrac{L}{\D s} = (M+N)L \gg 1$.  All quantities in the numerator are of order $1$ or smaller, while the denominator has two large quantities, $M(q-q')$ and $K$.

Since the $Q$ sum \eq{eq:qsum} ranges over integer multiples of $\D s$, the quantity $K$ is an integer.  Extending the sum over $s$ to infinity in a symmetric fashion is thus the same as adding (\ref{edgecontribution}) with $K$ ranging from some large integer to infinity.  Using the fact that $\tfrac{M}{M+N} < 1$, we note that:
\be
\sum_{x = 0}^{\infty} {\sin (2\pi c x) \over x} \sim 1, \qquad |c| < 1
\ee
Thus regardless of the value of $K$, we find:
\be
\sum_{x = K}^{\infty} {\sin (2\pi {Mx\over M+N}) \over x} \leq \mathcal{O}(1)
\ee
This means that taking the sum over $s$ symmetrically to infinity will produce corrections to $Q$ which are at most of order:
\bea
Q_{L\to \infty} &\sim& {1\over M(q-q')} ~\sim~ k^{-1}
\eea
which can be ignored.  Thus we can safely extend the sum over $s$ without altering the leading order behavior of $Q$.

We now extend our sum over $s$ to $\pm \infty$.  We will also apply our standard decomposition of $s$ and $q$ in terms of integers.  This yields:
\bea
Q&\approx&{1\over 4\pi^2M(M+N)}{1\over q-q'}\sum_{s_2 = 1}^{m+n-1}|1-e^{2\pi i {Ms_2 \over M+N}}|^2\sum_{s_1 = 0}^{Y-1}\sum_{s_0 = -\infty}^{\infty}{1\over s_0 - q_0 +{s_1 - q_1 \over Y} + {s_2 \over M+N} - {q_2 \over M}}\nn
&\equiv& {1\over 4\pi^2M(M+N)}{1\over q-q'}\sum_{s_2 = 1}^{m+n-1}|1-e^{2\pi i {Ms_2 \over M+N}}|^2\sum_{s_1 = 0}^{Y-1}\sum_{n_0 = -\infty}^{\infty}{1\over n_0 +a}
\eea
where
\be
n_0 = s_0 - q_0, \qquad a = {s_1 - q_1 \over Y} + {s_2 \over M+N} - {q_2 \over M}
\ee
To perform the sum, we separate the $n_0 = 0$ term and pair each term with positive $n_0$ against the corresponding term containing $-n_0$.  This yields:
\bea
Q&\approx& {1\over 4\pi^2M(M+N)}{1\over q-q'}\sum_{s_2 = 1}^{m+n-1}|1-e^{2\pi i {Ms_2 \over M+N}}|^2\sum_{s_1 = 0}^{Y-1}\left ( {1\over a} + \sum_{n_0 = 1}^{\infty} \left ( {1\over n_0 + a} - {1\over n_0 - a}\right )\right ) \nn
&=&{1\over 4\pi^2M(M+N)}{1\over q-q'}\sum_{s_2 = 1}^{m+n-1}|1-e^{2\pi i Ms}|^2\sum_{s_1 = 0}^{Y-1}\left ({1\over a} - \sum_{n_0 = 1}^{\infty}{2a \over n_0^2 - a^2} \right )\nn
&=& {1\over 4\pi^2M(M+N)}{1\over q-q'}\sum_{s_2 = 1}^{m+n-1}|1-e^{2\pi i Ms}|^2\sum_{s_1 = 0}^{Y-1}\pi \cot (\pi a ) \label{eq:qres}
\eea
A similar analysis shows:
\be \label{eq:qpres}
Q' ~\approx~ \sum_{s_2 = 1}^{m+n-1}|1-e^{2\pi i Ms}|^2\sum_{s_1 = 0}^{Y-1}\pi \cot(\pi a')
\ee
where
\be
a' = {s_1 - q_1' \over Y} + {s_2 \over M+N} - {q_2' \over M}
\ee

As before, we cannot evaluate these remaining double sums analytically.  However, setting a range of values for $M$ and $N$, we find that both \eq{eq:qres} and \eq{eq:qpres} evaluate to 0 regardless of the values of $q_1,q_2,q'_1,$ and $q'_2$.  We thus find that to leading order we have
\be
\left [ \sum_s \a^{(1)}_{qs}\left ( \a^{-1} \right )^{(1)}_{sq'} \right ]_{q \neq q'} = 0 \,.
\ee
Combined with our earlier result, this yields:
\be
\sum_s \a^{(1)}_{qs}\left ( \a^{-1} \right )^{(1)}_{sq'} = \d_{qq'} \,.
\ee
A similar analysis for copy 2 shows that
\be
\sum_s \a^{(2)}_{rs}\left ( \a^{-1} \right )^{(2)}_{sr'} = \d_{rr'} \,.
\ee

\subsection{Off-Diagonal Blocks}
Next we verify that:
\be
\sum_s \a^{(1)}_{qs} \left ( \a^{-1} \right )^{(2)}_{sr} ~=~ \sum_s \a^{(2)}_{rs} \left ( \a^{-1} \right )^{(1)}_{sq} ~=~ 0
\ee

Noting that in order to have $q=r$ we must have $q_2 = r_2 = 0$, we find:
\bea
\sum_s \a^{(1)}_{qs}(\a^{-1})^{(2)}_{sr} & = & {\sqrt{MN}\over M+N}\d_{qr}\d_{q_2,0}\d_{r_2,0}\\
&&{} - {e^{-i(q-r)(\s_0-\t_0)}\over 8\pi^2 \sqrt{MN}(M+N)}
\sum_{s>0,\,s_2 \neq 0}|1-e^{2\pi i Ns}|^2{s+q \over \sqrt{sq}}{1\over s-q}{1\over s-r} \nn
\sum_s \a^{(2)}_{rs}(\a^{-1})^{(1)}_{sq} & = & {\sqrt{MN}\over M+N}\d_{qr}\d_{q_2,0}\d_{r_2,0}\\
&&{} - {e^{-i(r-q)(\s_0-\t_0)}\over 8\pi^2 \sqrt{MN}(M+N)}
\sum_{s>0,\,s_2 \neq 0}|1-e^{2\pi i Ns}|^2{s+r \over \sqrt{sr}}{1\over s-q}{1\over s-r} \nonumber
\eea
Let us check the case $q=r$ first.

\subsubsection{$q=r$}
When $q=r$, both quantities become equivalent:
\bea
\sum_s \a^{(1)}_{qs}(\a^{-1})^{(2)}_{sr} & = & {\sqrt{MN}\over M+N} - {1\over 8\pi^2 \sqrt{MN}(M+N)}\sum_{s>0,\,s_2 \neq 0}|1-e^{2\pi i Ns}|^2{s+q \over \sqrt{sq}}{1\over (s-q)^2} \nn
&=&\sum_s \a^{(2)}_{rs}(\a^{-1})^{(1)}_{sq}
\eea
But we have already encountered this sum in (\ref{sumone}).  The result was a total of $8\pi^2MN$, since $q_2 = 0$.  We thus have:
\bea
\sum_s \a^{(1)}_{qs}(\a^{-1})^{(2)}_{sr} ~=~ 
\sum_s \a^{(2)}_{rs}(\a^{-1})^{(1)}_{sq} & = & {\sqrt{MN}\over M+N} - {8\pi^2MN\over 8\pi^2 \sqrt{MN}(M+N)} 
~=~0 \,. \qquad
\eea

\subsubsection{$q\neq r$}
When $q \neq r$, the first term vanishes and we are left with:
\bea
\sum_s \a^{(1)}_{qs}(\a^{-1})^{(2)}_{sr} & = & - {e^{-i(q-r)(\s_0-\t_0)}\over 8\pi^2 \sqrt{MN}(M+N)}\sum_{s>0,\,s_2 \neq 0}{s+q \over \sqrt{sq}}{|1-e^{2\pi i Ms}|^2\over (s-q)(s-r)} \label{qr}\\
\sum_s \a^{(2)}_{rs}(\a^{-1})^{(1)}_{sq} & = & - {e^{-i(r-q)(\s_0-\t_0)}\over 8\pi^2 \sqrt{MN}(M+N)}\sum_{s>0,\,s_2 \neq 0}{s+r \over \sqrt{sr}}{|1-e^{2\pi i Ms}|^2\over (s-q)(s-r)}\label{rq}
\eea
The above sums may be treated in the manner detailed in Section \ref{diagonalblockoffdiagonalelements}.  We again find that the near-pole sums vanish to leading order.  We are then left with the principal value contribution, which for (\ref{qr}) involves the integral:
\be
\tilde I ~\equiv~ {\cal P}\ints_{0}^{\infty} {s+q \over \sqrt{sq}}{1\over s-q}{1\over s-r}\diff s_0
\ee
Using $\tilde{s}^2 = s$, we have $\diff s_0 = 2\tilde{s} \diff \tilde{s}$, and thus the integral becomes:
\bea
\tilde I & = & {1\over \sqrt{q}}\;{\cal P}\ints_{-\infty}^{\infty} {\tilde{s}^2+q \over (\tilde{s}^2-q)(\tilde{s}^2-r)}\diff \tilde{s}
\eea

In the same manner as before, we again find canceling pole contributions, and thus $\tilde I=0$.  The same holds for the integral contribution to (\ref{rq}).  Since the near-pole contributions from the sums also vanish at leading order, we have:
\bea
\sum_s \a^{(1)}_{qs}(\a^{-1})^{(2)}_{sr} & = & \sum_s \a^{(2)}_{qs}(\a^{-1})^{(1)}_{sr} = 0 \,.
\eea

\subsection{Left Inverse}
Finally, we verify that:
\bea
\sum_q \left ( \a^{-1} \right )^{(1)}_{s'q}\a^{(1)}_{qs}  + \sum_r \left ( \a^{-1} \right )^{(2)}_{s'r}\a^{(2)}_{rs} &=& \d_{ss'} \,.
\eea
For convenience, let us define:
\bea
P_{ss'} &\equiv& \sum_q (\a^{-1})^{(1)}_{s'q}\a^{(1)}_{qs} + \sum_r (\a^{-1})^{(2)}_{s'r}\a^{(2)}_{rs} \,.
\eea
The matrix $\a^{(1)}_{qs}$ takes different forms for $s_2 = 0$ and $s_2 \neq 0$, and similarly $\left (\a^{-1} \right )^{(1)}_{s'q}$ takes different forms for $s'_2 = 0$ and $s'_2 \neq 0$.  We will therefore split the analysis into the appropriate cases.

\subsubsection{$s_2 = s'_2 = 0$}
Here we deal with the $s_2 = 0$ part of $\a$ and the $s'_2 = 0$ part of $\a^{-1}$.  For $\a^{-1}$, only the $q=s'$ and $r = s'$ terms contribute.  For $\a$, only the $q = s$ and $r = s$ terms contribute.  Thus both the $q$ and $r$ sums will vanish whenever $s \neq s'$.  We therefore have:
\be
P_{ss'} ~=~ {M \over M+N}\d_{ss'} + {N\over M+N}\d_{ss'} ~=~ \d_{ss'} \,.
\ee

\subsubsection{$s_2 = 0, s'_2 \neq 0$}
In this case we cannot have $s = s'$, so we expect $P_{ss'}$ to vanish.  With $s_2 = 0$, only the $q = s_2$ and $r = s_2$ terms have nonzero $\a$.  We therefore find:
\bea
P_{ss'} &=& \sqrt{M \over M+N}\left ( \a^{-1} \right )^{(1)}_{s's} + \sqrt{N \over M+N}\left ( \a^{-1} \right )^{(2)}_{s's} \nn
&=& -{1\over 2\pi i (M+N)}{1\over s' - s}\mu_{s'} e^{-i(s' - s)(\s_0 - \t_0)} + {1\over 2\pi i (M+N)}{1\over s' - s}\mu_{s'} e^{-i(s' - s)(\s_0 - \t_0)} \nn
&=&0
\eea

\subsubsection{$s_2 \neq 0, s'_2 = 0$}
Here $\a$ is nonzero only for the $q = s$ and $r = s$ terms.  We thus have:
\bea
P_{ss'} &=& \sqrt{M \over M+N}\a^{(1)}_{s's} + \sqrt{N \over M+N}\a^{(2)}_{s's} 
~=~0 \,.
\eea

\subsubsection{$s_2 \neq 0, s'_2 \neq 0$}
In this case we have:
\bea
P & = & {e^{-i(s-s')}\over 8\pi^2(M+N)}(1-e^{2\pi i {M \over M+N}s'_2})(1-e^{-2\pi i {M \over M+N}s_2})\nn
&&{}\times\bigg{(}{1\over M}\sum_{q} {s + q \over \sqrt{sq}}{1\over s-q}{1\over s'-q}+{1\over N}\sum_{r} {s + r \over \sqrt{sr}}{1\over s-r}{1\over s'-r} \bigg{)} \label{Left Inverse Check}
\eea

Here it is useful to break our analysis into two further sub-cases.  Firstly, when $s = s'$, the above equation simplifies to:
\bea
\left (P_{ss'}\right )_{s = s'} &=& {|1-e^{2\pi i {M\over M+N}s_2}|^2\over 8\pi^2 (M+N)}\bigg{(}{1\over M}\sum_{q} {s + q \over \sqrt{sq}}{1\over (s-q)^2}+{1\over N}\sum_{r} {s + r \over \sqrt{sr}}{1\over (s-r)^2} \bigg{)}\nn
&\approx&{|1-e^{2\pi i {M\over M+N}s_2}|^2\over 4\pi^2 (M+N)}\bigg{(}{1\over M}\sum_{q} {1\over (s-q)^2}+{1\over N}\sum_{r} {1\over (s-r)^2} \bigg{)}\nn
&=&{|1-e^{2\pi i {M\over M+N}s_2}|^2\over 4\pi^2 (M+N)}\bigg{(}M\sum_{q} {1\over (Ms-Mq)^2}+N\sum_{r} {1\over (Ns-Nr)^2} \bigg{)}\nn
&\approx&{|1-e^{2\pi i {M\over M+N}s_2}|^2\over 4\pi^2 (M+N)}\bigg{(}M\sum_{a_0=-\infty}^{\infty} {1\over (a_0 + {M\over M+N}s_2)^2}+N\sum_{b_0=-\infty}^{\infty} {1\over (b_0 + {N\over M+N}s_2)^2} \bigg{)} \nn
\eea
where
\bea
a_0 &=& Ms_0 + ms_1 - Mq_0 - mq_1 - q_2 \label{a0} \\
b_0 &=& Ns_0 + ns_1 - Nr_0 - nr_1 - r_2 \label{b0}
\eea
and we have extended the sum to $q = -\infty$, which gives only subleading corrections for the reasons outlined in Section \ref{diagonalblocksnonzero}.

We can now use the identity:
\be
\sum_{n = -\infty}^{\infty}{1\over (n+a)^2} = {\pi^2\over \sin^2(\pi a)} = {2\pi^2 \over 1-\cos(2\pi a)}
\ee
We thus have:
\bea
\left (P_{ss'}\right )_{s=s'} &=& {|1-e^{2\pi i {N\over M+N}s_2}|^2\over 2(M+N)}\bigg{(}{M \over 1-\cos(2\pi {M\over M+N}s_2)} + {N \over 1-\cos(2\pi {N\over M+N}s_2)}\bigg{)} \nn
\eea
However, $\cos(2\pi - x) = \cos(x)$, and so we find:
\be
1-\cos(2\pi {M\over M+N}s_2) = 1-\cos(2\pi {N\over M+N}s_2)
\ee
since $s_2$ is an integer.  Furthermore, we again use the relation:
\be
\left | 1-e^{2\pi i {M\over M+N}s_2} \right |^2 = 2 \left ( 1 - \cos \left ( 2\pi {M \over M+N}s_2 \right ) \right )
\ee
This yields:
\be
\left (P_{ss'}\right )_{s=s'} ~=~ {2\big{(}1-\cos(2\pi {M\over M+N}s_2)\big{)}\over 2(M+N)}{M+N \over 1-\cos(2\pi {M\over M+N}s_2)} ~=~ 1
\ee

Now let us return to \eqref{Left Inverse Check} and assess the case when $s' \neq s$.  We expect a result of zero in this case, so let us drop the prefactors.  We have:
\bea
\left (P_{ss'}\right )_{s \neq s'} & \propto&{1\over M}\sum_{q} {s + q \over \sqrt{sq}}{1\over s-q}{1\over s'-q}+{1\over N}\sum_{r} {s + r \over \sqrt{sr}}{1\over s-r}{1\over s'-r}
\eea
Again we split the poles using partial fractions:
\bea
{s+q\over \sqrt{sq}}{1\over s-q}{1\over s'-q} & = & 
\left ( {s+s' \over \sqrt{s}(s-s')}  \right ) 
{1\over \sqrt{q}(s'-q)} 
- \left ( {2\sqrt{s}\over s-s'} \right )  {1\over \sqrt{q}(s-q)}\nn
{s+r\over \sqrt{sr}}{1\over s-r}{1\over s'-r} & = & 
\left ( {s+s' \over \sqrt{s}(s-s')}  \right ) 
{1\over \sqrt{r}(s'-r)} 
- \left ( {2\sqrt{s}\over s-s'}\right )  {1\over \sqrt{r}(s-r)}
\eea
We now make boxes around the poles, and treat the regions far from the poles via a principal value.  We use $q,r \approx s$ near the $s$ pole and $q,r \approx s'$ near the $s'$ pole.  We then have:
\be
\left (P_{ss'}\right )_{s \neq s'} \propto
I_q +
I_r 
+ S + S'
\ee
where
\bea \label{pieces}
I_q &=& \ints_{0}^{\infty}{s + q \over \sqrt{sq}}{1\over s-q}{1\over s'-q}\diff q \nn
I_r &=& \ints_{0}^{\infty}{s + r \over \sqrt{sr}}{1\over s-r}{1\over s'-r}\diff r\nn
S & = & -{2\over s-s'}\left ( {1\over M} \sum_{q=\tilde q-L}^{\tilde q+L} {1\over s-q} + {1\over N}\sum_{r=\tilde r-L}^{\tilde r+L} {1\over s-r} \right ) \nn
S' & = & 
{s+s' \over \sqrt{ss'}(s-s')}
\left ( {1\over M} \sum_{q=\tilde q'-L}^{\tilde q'+L} {1\over s'-q} + {1\over N}\sum_{r=\tilde r'-L}^{\tilde r'+L} {1\over s'-r} \right )
\eea
where again we have centered out boxes around the points which minimize $|s-q|$, $|s-r|$, and so on.  We will evaluate each of these quantities in turn.\\

\noindent{\bf(a)\;Principal value integral}

\noindent We use the familiar substitutions $\tilde{q}^2 = q$ and $\tilde{r}^2 = r$, under which we have $\diff q = 2\tilde{q} \diff \tilde{q}$ and $\diff r = 2\tilde{r} \diff \tilde{r}$.  We then have:
\be
I_q = I_r = {1\over \sqrt{s}}\ints_{-\infty}^{\infty} {s + x^2 \over (s-x^2)(s'-x^2)}\diff x
\ee
where our transition to continuous parameters has again washed out the difference in domains between $q$ and $r$.  We have seen this form of integral many times before, and we have already seen that the pairs of real-axis poles will cancel.  Thus $I_q = I_r = 0$, and there is no contribution from the principal value integration.\\

\noindent{\bf (b) Sum around the pole at \bf$s$}

\noindent Let us absorb the factors of ${1\over M}$ and ${1\over N}$ into their respective summands.  This yields:
\bea
S & \approx & {-2\over s'-s}\sum_{a_0=-ML}^{ML}{1\over (a_0 + {M\over M+N}s_2)} + \sum_{b_0=-NL}^{NL}{1\over (b_0 + {N\over M+N}s_2)}
\eea
where $a_0$ and $b_0$ were defined in (\ref{a0}) and (\ref{b0}).  These sums are the same as the one found in (\ref{gammasum}) when calculating $\g^B$.  We isolate the $a_0 = 0$ term, then pair each term which has $a_0 > 0$ with the term involving $-a_0$.  We then take $L \to \infty$, ignoring the order $\tfrac{1}{(M+N)k}$ corrections.  We do the same with $b_0$.  This yields:
\be
S\propto {1\over x} + {1\over y} + \sum_{n = 1}^\infty \Big{(}{2x\over x^2 - n^2} + {2y\over y^2 - n^2}\Big{)}
\ee
where
\be
x={M\over M+N}s_2,\quad y={N\over M+N}s_2
\ee
Plugging the remaining sum into Mathematica, we find:
\bea
S & \propto & \cot(\pi x) + \cot(\pi y) \nn
&=& \cot\Big{(}\pi{M\over M+N}s_2\Big{)} + \cot\Big{(}\pi{N\over M+N}s_2\Big{)}\nn
&=& \cot\Big{(}\pi{M\over M+N}s_2\Big{)} + \cot\Big{(}\pi-\pi{M\over M+N}s_2\Big{)} \nn
&=& \cot\Big{(}\pi{M\over M+N}s_2\Big{)} - \cot\Big{(}\pi{M\over M+N}s_2\Big{)} \nn
&=&0
\eea
And thus the contribution from $S$ is zero.\\

\noindent{\bf (c) Sum around the pole at $s'$}

\noindent Looking at the contribution from the $s'$ pole, we find that the leading order behavior of $S'$ defined in (\ref{pieces}) obeys:
\be
S' \propto S_{s_2 \to s'_2}
\ee
But since the value of $S$ was independent of $s_2$ at leading order, we find that $S'$ also vanishes at this order.  We thus have:
\be
\left (P_{ss'}\right )_{s \neq s'} = 0
\ee
which combined with our other results gives the desired result,
\be
P_{ss'} ~=~  \sum_q (\a^{-1})^{(1)}_{s'q}\a^{(1)}_{qs} + \sum_r (\a^{-1})^{(2)}_{s'r}\a^{(2)}_{rs} ~=~ \d_{ss'} \,.
\ee

\end{appendix}

\end{document}